\documentclass[reprint,amsmath,amssymb,aps,prx,floatfix,longbibliography,superscriptaddress
]{revtex4-2}

\usepackage{graphicx}
\usepackage{bm}
\usepackage{booktabs}
\usepackage{mathtools}
\usepackage{placeins}
\usepackage[svgnames]{xcolor}
\usepackage[bookmarks=false,linkcolor=blue,urlcolor=blue,colorlinks,citecolor=blue]{hyperref}
\usepackage{tikz}
\usetikzlibrary{arrows.meta,positioning}

\newcommand{\ii}{\mathrm{i}}
\newcommand{\softmax}{\operatorname{softmax}}


\begin{document}

\title{Learning Quantum Matter through Attention in Complex Space}

\author{Mingrui Jing}
\thanks{These authors contributed equally to this work.}
\affiliation{QudeLeap Research, Shanghai 200030, China}
\affiliation{Thrust of Artificial Intelligence, Information Hub,\\
The Hong Kong University of Science and Technology (Guangzhou), Guangzhou 511453, China}
\author{Erdong Huang}
\thanks{These authors contributed equally to this work.}
\affiliation{QudeLeap Research, Shanghai 200030, China}
\affiliation{Thrust of Artificial Intelligence, Information Hub,\\
The Hong Kong University of Science and Technology (Guangzhou), Guangzhou 511453, China}
\author{Jizhe Lai}
\author{Enji Xiong}
\affiliation{Thrust of Artificial Intelligence, Information Hub,\\
The Hong Kong University of Science and Technology (Guangzhou), Guangzhou 511453, China}
\author{Jin-Guo Liu}
\thanks{jinguoliu@hkust-gz.edu.cn}
\affiliation{Advanced Materials Thrust, Function Hub,\\
The Hong Kong University of Science and Technology (Guangzhou), Guangzhou 511453, China}
\author{Xin Wang}
\thanks{felixxinwang@hkust-gz.edu.cn}
\affiliation{Thrust of Artificial Intelligence, Information Hub,\\
The Hong Kong University of Science and Technology (Guangzhou), Guangzhou 511453, China}

\begin{abstract}
Magnetic many-electron wavefunctions require amplitude and phase to
be optimized together. Whether a complex internal representation improves
this variational search is a practical question for neural wavefunction design.
We introduce Complex Psiformer for interacting electrons in a magnetic
moir\'e continuum, combining complex hidden features and Hermitian-magnitude
attention with magnetic boundary conditions and fermionic antisymmetry.
After the same number of optimization steps, Complex Psiformer reaches
lower energies than Real Psiformer in two finite supercells.
Both Psiformers also improve on their respective neural Hartree--Fock references.
Across five training seeds in the $25$-cell system, the mean Complex
advantage is $1.458\,\mathrm{meV}$ per electron, with a smaller observed spread.
A separately trained two-electron Complex state has a smaller energy gap
to a finite configuration-interaction reference than its Real counterpart.
In the Complex states, flux scans show nonmonotonic density correlations
and weaker honeycomb mean-density modulation at higher flux, while connected
fluctuations persist.
Gauge-invariant current maps provide a qualitative comparison of local
circulation in the optimized states.
These benchmarks support the combined architecture as a variational ansatz
for studying energies and charge arrangements in finite magnetic systems.
\end{abstract}

\maketitle

\section{Introduction}
\label{sec:introduction}

Variational descriptions of interacting electrons in a perpendicular magnetic
field must account for strong electronic correlations and the configuration-dependent
phase of the many-body wavefunction. In a periodic supercell, fermionic
antisymmetry must be combined with gauge-dependent boundary transition phases~\cite{haldane1985translational,onofri2001landau}. Enforcing these boundary
conditions leaves the many-body amplitude and phase within the cell to be
optimized. The phase gradients contribute to the covariant kinetic energy
and determine local currents together with the amplitude and vector potential.
Moir\'e superlattices bring this problem to experimentally accessible fields,
where commensurate charge order and Hofstadter states compete as the flux per
moir\'e cell varies~\cite{dean2013hofstadter,kometter2023hofstadter}.

First-quantized fermionic wavefunctions conventionally enforce antisymmetry
through Slater determinants and describe correlations with Jastrow factors,
backflow transformations, or multideterminant expansions~\cite{foulkes2001qmc,luo2019backflow}. Neural quantum states introduce
trainable representations optimized by variational Monte Carlo (VMC)~\cite{carleo2017solving,choo2020fermionic}. Early studies established the use of neural-network representations
for fermionic many-body states~\cite{cai2018approximating}. SlaterNet learns one-electron orbitals, while
FermiNet and PauliNet construct orbitals that depend on the electronic
configuration~\cite{pfau2020ferminet,hermann2020paulinet}. Attention-based
architectures introduce this dependence by exchanging information between
electron streams before determinant projection~\cite{geier2025attention,vonglehn2023self}. These approaches have produced
accurate results for molecules, periodic solids, moir\'e materials, and quantum
Hall systems~\cite{pescia2022neuralnetwork,li2022deepsolid,wilson2023wapnet,teng2025solving}.

Several first-quantized ansatzes describe complex wavefunctions using real
hidden features together with complex orbital readouts or additional
complex-valued factors~\cite{geier2025attention,teng2025solving,perezfadon2025anyon,abouelkomsan2026magnet}.
In the Real Psiformer baseline considered here, attention exchanges
real-valued features before the orbital readout forms complex coefficients.
Organizing hidden electron features as complex amplitudes gives
radial nonlinearities and Hermitian overlaps a direct role in information
exchange before orbital construction. Can the complex structure of quantum
mechanics guide a more effective variational search for magnetic
many-electron states from first principles?

In this manuscript, we introduce \emph{Complex Psiformer} for
interacting electrons in a magnetic moir\'e continuum. Here, first-principles
means direct variational optimization of the specified effective continuum
Hamiltonian. Each orbital combines a fixed section carrying the magnetic
boundary transition law with a learned periodic configuration-dependent
coefficient. The coefficient network uses reciprocal-lattice phasors,
signed radial activations, and Hermitian-magnitude complex-value attention
(HMCVA). We adapt this complex attention pattern to electron
streams~\cite{eilers2023building}, using real attention weights to aggregate
complex-valued messages. The resulting orbitals form Slater determinants
that enforce fermionic antisymmetry. We also define \emph{Complex SlaterNet}
as an attention-free reference with the same complex orbital construction.
The main comparisons use minimum-step stochastic reconfiguration
(MinSR)~\cite{chen2024minsr,rende2024sridentity} for variational optimization.

We evaluate the complete Complex Psiformer architecture against Real Psiformer and real and complex
neural Hartree--Fock (HF) baselines in two finite magnetic supercells.
Both Psiformers lower the energies relative to their respective neural HF
baselines. After the same number of optimization steps, Complex Psiformer further
lowers the mean energy by $1.46\,\mathrm{meV}$ per electron relative to
Real Psiformer over all five attempted training seeds in the $25$-cell system.
A separately trained $N=2$ Complex state also has a smaller energy
gap to a finite configuration-interaction reference than its Real
counterpart. In the $18$-cell system, the clearer
honeycomb-type charge pattern of Complex Psiformer at $n_\phi=2$ is reflected
mainly in the mean-density contribution to the structure factor. A flux scan
reveals a nonmonotonic redistribution of density correlations.
Gauge-invariant current maps show the accompanying local circulation
qualitatively.

\section{Magnetic moir\'e model and neural wavefunctions}
\label{sec:model-and-wavefunctions}

The continuum Hamiltonian and magnetic boundary conditions define the
finite-supercell problem. Complex one-particle orbitals form Complex SlaterNet,
and attention between electron streams extends it to Complex Psiformer.
Section~\ref{sec:vmc} describes the VMC sampling and optimization procedure.

\subsection{Magnetic moir\'e continuum}
\label{sec:system}

We consider an interacting two-dimensional electron gas in a triangular moir\'e potential under a uniform perpendicular magnetic field. At zero field, the Hamiltonian reduces to the standard effective-mass continuum model for spin-polarized carriers in two-dimensional moir\'e semiconductors
\cite{wu2018hubbard,zhang2020moirequantumchemistry,luo2024neuralbloch,
geier2025attention}. This model describes aligned transition-metal
dichalcogenide (TMD) heterobilayers such as
$\mathrm{WSe}_2/\mathrm{WS}_2$, in which Mott and generalized Wigner-crystal
states have been observed at integer and fractional moir\'e fillings
\cite{regan2020mott}. We work directly in the continuum
without projecting onto moir\'e bands or Landau levels.

\begin{figure}[h]
    \centering
    \includegraphics[width=0.9\linewidth]{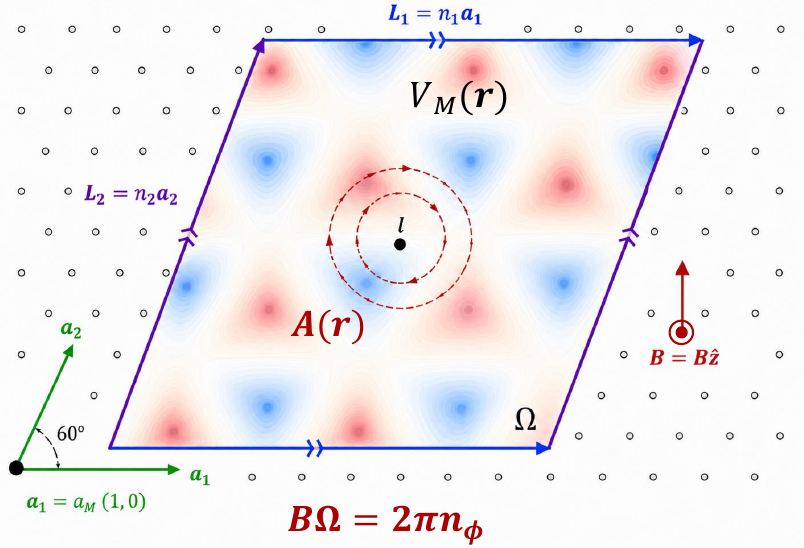}
    \caption{Geometry of the triangular moiré system. The real-space moiré lattice has primitive vectors $\bm{a}_{1,2}$ and an $n_1\times n_2$ periodic supercell spanned by $\bm{L}_{1,2}$, with area $\Omega$. The color map represents the moiré potential. A uniform field $\bm{B}=B\hat{z}$ is applied perpendicular to the plane, and the symmetric gauge is centered at $l$. Matching edge markers denote periodic identification of opposite boundaries.}
    \label{fig:system}
\end{figure}

The triangular moir\'e lattice is generated by $\bm a_1=a_{\mathrm M}(1,0)$ and $\bm a_2=a_{\mathrm M}(1/2,\sqrt{3}/2)$, where $a_{\mathrm M}$ is the moir\'e lattice constant. The reciprocal vectors $\bm g_1$ and $\bm g_2$ are defined by $\bm g_a\cdot\bm a_b=2\pi\delta_{ab}$, and $\bm g_3=-(\bm g_1+\bm g_2)$. The three $\bm g_a$ are related by $120^\circ$ rotations. Together with their negatives they form the first star of the moir\'e reciprocal lattice. Keeping this first star gives the standard leading-harmonic potential 
\begin{equation} V_{\mathrm M}(\bm r) =-2V_0\sum_{a=1}^{3}\cos(\bm g_a\cdot\bm r+\varphi), \label{eq:moire-potential} \end{equation} 
used in continuum descriptions of triangular TMD moir\'e systems \cite{wu2018hubbard,zhang2020moirequantumchemistry,geier2025attention}. Reality and threefold rotation symmetry reduce the first-star Fourier coefficients to a common amplitude $V_0$ and a single moir\'e phase $\varphi$. This phase sets the relative energies of high-symmetry stacking regions and thereby shapes the potential landscape. It is unrelated to the phase of the variational wavefunction. Calculations are performed in an $n_1\times n_2$ periodic supercell, with $n_1,n_2\in\mathbb N_{>0}$, $\bm L_1=n_1\bm a_1$, $\bm L_2=n_2\bm a_2$, and area $\Omega=|\bm L_1\times\bm L_2|>0$. The corresponding reciprocal vectors are $\bm G_1=\bm g_1/n_1$ and $\bm G_2=\bm g_2/n_2$, and hence $\bm G_a\cdot\bm L_b=2\pi\delta_{ab}$. The vectors $\bm g_a$ define the physical moir\'e potential, whereas $\bm G_a$ define features that are periodic over the finite simulation cell. We now apply a uniform field $\bm B=B\hat{\bm z}$ perpendicular to the plane. Identifying opposite edges of the finite cell makes its geometry a torus, so a consistent magnetic boundary condition requires an integer number of flux quanta through the cell. In the effective atomic units used here,
\begin{equation} B\Omega=2\pi n_{\phi}, \quad n_{\phi}\in\mathbb Z_{\geq0}. \label{eq:flux-quantization} \end{equation}
Thus $n_{\phi}$ is the number of flux quanta through the supercell. At
$n_{\phi}=0$, the field vanishes, the magnetic section is not applied, and the
wavefunction obeys ordinary periodic boundary conditions. The system is illustrated in Figure~\ref{fig:system}.

We
introduce the field through minimal coupling to a vector potential satisfying
$(\bm\nabla\times\bm A)_z=B$. We use the symmetric gauge centered at
$\bm l=(l_x,l_y)=(\bm L_1+\bm L_2)/2$, for which
$\bm A(\bm r)=B(-(y-l_y),x-l_x)/2$. Choosing the center of the finite cell as
$\bm l$ is convenient but does not change the magnetic field or any
observable. The carrier-charge sign is absorbed into the charge-weighted
vector potential, and the finite-field calculations use $B>0$ in the
convention of Equation~\eqref{eq:hamiltonian}.

Writing the Hamiltonian in effective atomic units, whose material-dependent conversion is given at the beginning of Sec.~\ref{sec:main-results}, we obtain
\begin{equation} \begin{multlined} \hat H=\frac{1}{2}\sum_{i=1}^{N} \bigl(-\ii\bm\nabla_i-\bm A(\bm r_i)\bigr)^2 +\sum_{i=1}^{N}V_{\mathrm M}(\bm r_i)\\ +\sum_{i<j}v_{\mathrm{Ew}}(\bm r_i-\bm r_j) +\frac{N}{2}\xi_{\mathrm M}. \end{multlined} \label{eq:hamiltonian} 
\end{equation} 
Here $N$ is the number of electrons, and $\bm\nabla_i$ differentiates
with respect to the coordinate $\bm r_i$. The first two terms describe the
minimally coupled kinetic energy and the moir\'e potential. We evaluate the
Coulomb interaction in the periodically repeated supercell using the
two-dimensional Ewald form. The periodic pair interaction $v_{\mathrm{Ew}}$
includes the neutralizing-background correction, while $\xi_{\mathrm M}$ is the Madelung self-energy. Together, these contributions give the full Coulomb energy~\cite{fraser1996finitesize,foulkes2001qmc,geier2025attention}. Their explicit forms are given in Appendix~\ref{app:ewald}.

The symmetric-gauge vector potential is not periodic. Under a supercell translation, it changes by the gauge transformation $\bm A(\bm r+\bm L_a)=\bm A(\bm r)+\bm\nabla\Lambda_a(\bm r)$, where $\Lambda_a(\bm r)=B[\bm L_a\times(\bm r-\bm l)]_z/2$. Gauge covariance of the kinetic energy therefore fixes the magnetic boundary condition.
\begin{equation}
\Psi(\ldots,\bm r_i+\bm L_a,\ldots)
=e^{\ii\Lambda_a(\bm r_i)}\Psi(\ldots,\bm r_i,\ldots).
\label{eq:magnetic-boundary}
\end{equation} 
The vector potential fixes the gauge-transition phase $e^{\ii\Lambda_a(\bm r)}$.
It contains no variational parameter or additional boundary twist. Translating the same electron by $\bm L_1$ and
$\bm L_2$ in opposite orders yields the relative phase $e^{-\ii B\Omega}$.
Equation~\eqref{eq:flux-quantization} sets this phase to unity and ensures that
the two magnetic boundary identifications are consistent
\cite{onofri2001landau,haldane1985translational}. In the neural wavefunctions
below, the learned coefficients remain periodic over the supercell. Fixed
single-particle functions supply the magnetic translation law in
Equation~\eqref{eq:magnetic-boundary}.

\subsection{Neural-wavefunction hierarchy}
\label{sec:neural-wavefunctions}

\begin{figure}[t]
\centering
\includegraphics[width=0.9\linewidth]
{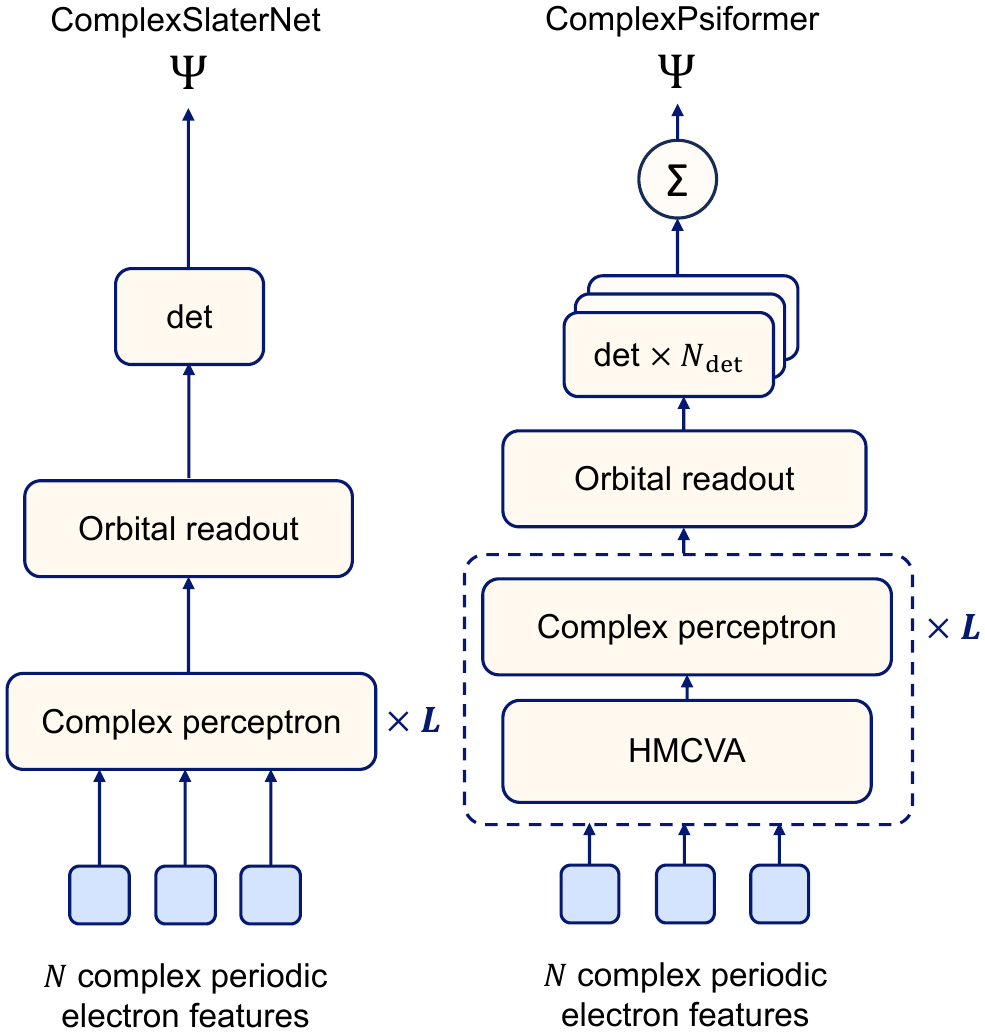}
\caption{Construction of the two complex-valued neural wavefunctions.  Complex
SlaterNet applies the same one-body network independently to each electron.
Complex Psiformer inserts HMCVA before every residual feed-forward layer.  This
allows each electron representation to incorporate information from the others
before the orbital readout and Slater determinant.  Both models use the same
fixed functions at the orbital readout to enforce the magnetic boundary
condition.  These functions are omitted from the schematic.}
\label{fig:model-comparison}
\end{figure}

Following Ref.~\cite{geier2025attention}, we combine a periodic neural
coefficient with the magnetic boundary structure of a single-particle orbital. Applying this construction
independently to every electron gives Complex SlaterNet. HMCVA then couples the
electron representations and produces the configuration-dependent orbitals of
Complex Psiformer. Figure~\ref{fig:model-comparison} summarizes both
architectures. Appendix~\ref{app:variational-review} places them in the broader
context of variational fermionic wavefunctions.

\subsection{From plane waves to one-body orbitals}
\label{sec:one-body-orbitals}

We construct each neural orbital from a coordinate feature, a
feed-forward network, and a scalar readout. The first step maps the coordinate
to a complex feature vector. For a periodic supercell, we choose a finite set
$\{\bm K_\mu\}_{\mu=1}^{d_f}$ of supercell reciprocal-lattice vectors. Each
vector satisfies $\bm K_\mu\cdot\bm L\in2\pi\mathbb Z$ for every supercell
translation $\bm L$. The feature map is
\begin{equation}
\bm f^0(\bm r)
\equiv \operatorname{feature}(\bm r)
=
\left(e^{\ii\bm K_1\cdot\bm r},\ldots,
      e^{\ii\bm K_{d_f}\cdot\bm r}\right)\in\mathbb C^{d_f},
\label{eq:complex-features}
\end{equation}
where $d_f$ is the number of selected modes and need not equal the
spatial dimension $d$. The set may contain fundamental modes, higher
harmonics, or other reciprocal-lattice combinations. Every component of
$\bm f^0$ is continuous and periodic under supercell translations. Complex
phasors contain the same information as the paired sine and cosine coordinates
used in
Refs.~\cite{pescia2022neuralnetwork,cassella2023discovering,geier2025attention}.
These inputs respect supercell translation symmetry.
Each $e^{\ii\bm K_\mu\cdot\bm r}$ is a plane-wave basis function for periodic
functions. The nonlinear network nevertheless uses the selected modes as
features and does not restrict the orbital to their linear span. Our
calculations use the two-dimensional specialization
${\bm K_1,\bm K_2}={\bm G_1,\bm G_2}$ and $d_f=2$. The phasor arguments encode
periodic position rather than the phase of the many-body wavefunction.

Second, a bias-free embedding maps the coordinate feature to the first hidden
representation,
\begin{equation}
\bm h^0(\bm r)
=W^0\bm f^0(\bm r)+V^0\overline{\bm f^0(\bm r)} .
\label{eq:complex-embedding}
\end{equation}
Here $d_{\rm L}$ is the common hidden-layer width, and
$W^0,V^0\in\mathbb C^{d_{\rm L}\times d_f}$ map the input to
$\bm h^0\in\mathbb C^{d_{\rm L}}$. The overline denotes componentwise complex
conjugation. Because the orbital is a complex-valued function of real
coordinates, the embedding need not be holomorphic. The term $W^0\bm f^0$
alone defines a complex-linear map. Including
$V^0\overline{\bm f^0}$ gives a general real-linear map between the underlying
real vector spaces. For the phasor features in
Equation~\eqref{eq:complex-features}, the conjugate channel also provides the
opposite reciprocal modes $e^{-\ii\bm K_\mu\cdot\bm r}$.

The network contains $L$ residual layers. For $l=0,\ldots,L-1$, layer $l+1$
uses the real-affine map
\begin{equation}
\mathcal F^{l+1}(\bm z)
=W^{l+1}\bm z+V^{l+1}\overline{\bm z}+\bm b^{l+1} .
\label{eq:free-real-linear}
\end{equation}
Here $W^{l+1},V^{l+1}\in
\mathbb C^{d_{\rm L}\times d_{\rm L}}$ and
$\bm b^{l+1}\in\mathbb C^{d_{\rm L}}$. The linear part is real-linear, while
the bias makes the complete map real-affine. We then apply a signed radial
activation to $\bm z\in\mathbb C^{d_{\rm L}}$.
\begin{equation}
\sigma^{l+1}(\bm z)
=\tanh\left(|\bm z|-\bm\Delta^{l+1}\right)
\frac{\bm z}{|\bm z|} .
\label{eq:radial-activation}
\end{equation}
This activation follows amplitude--phase constructions used in
complex-valued networks, including the pointwise modReLU activation
\cite{hirose1992continuous,arjovsky2016unitary}.
We use a signed $\tanh$ response. In numerical calculations, we
replace $|\bm z|$ in the denominator by $|\bm z|+10^{-8}$ to regularize
the activation at the origin. All operations in
Equation~\eqref{eq:radial-activation} act elementwise. The trainable offsets
$\bm\Delta^{l+1}\in\mathbb R^{d_{\rm L}}$, initialized at zero, shift the
crossover of the radial response.
A negative radial response reverses the channel phase by $\pi$.
The regularity relevant to energy evaluation is discussed in
Appendix~\ref{app:complex-vmc}. The complete residual update is
$\bm h^{l+1}=\bm h^l+
\sigma^{l+1}(\mathcal F^{l+1}(\bm h^l))$.

After $L$ residual layers, a trainable vector
$\bm w\in\mathbb C^{d_{\rm L}}$ produces the periodic neural coefficient
$\bm w\cdot\bm h^L(\bm r)$. At zero field, this coefficient can serve directly
as a periodic orbital. In a magnetic field, the orbital must instead acquire
the position-dependent transition phase $e^{\ii\Lambda_a(\bm r)}$ in
Equation~\eqref{eq:magnetic-boundary} under a supercell translation. We
separate this prescribed boundary law from the trainable periodic coefficient
by introducing fixed single-particle functions $\chi_j$ satisfying
\begin{equation}
\chi_j(\bm r+\bm L_a)=e^{\ii\Lambda_a(\bm r)}\chi_j(\bm r) .
\label{eq:magnetic-orbital-factorization}
\end{equation}
The index $j$ labels an orbital column in the Slater matrix. In the
language of magnetic translations, each $\chi_j$ is a magnetic section
\cite{onofri2001landau}. It is represented within one cell, with opposite
edges related by the required gauge-transition phases. These fixed, nontrainable functions contain amplitude and phase structure
within the cell and obey the same transition law. Multiplying a periodic neural
coefficient by $\chi_j$ therefore gives an orbital with the correct magnetic
boundary condition. The construction of $\chi_j$ uses torus Landau-type
functions. Multiplication by an arbitrary periodic neural coefficient does
not, however, restrict the orbital to a fixed Landau-level subspace.
Each orbital inherits the zeros of its fixed factor.
A zero of one orbital at one electron position need not make the determinant
vanish, so these factors alone do not determine the nodes of the full
many-electron wavefunction.
Appendix~\ref{app:magnetic-sections} describes the construction of $\chi_j$.
Every model considered below uses this same factorization, with the
fixed magnetic section carrying the prescribed boundary phase factor.

\subsection{Complex SlaterNet}
\label{sec:complex-slater}

Complex SlaterNet constructs an antisymmetric many-electron state by
applying the one-body network independently to each electron. Electron $i$
starts from $\bm f_i^0=\bm f^0(\bm r_i)$ and passes through the single-particle
stream $\bm h_i^0,\ldots,\bm h_i^L$ described above
\cite{geier2025attention}. All electrons share the embedding and
residual-layer parameters. Since the layers do not mix electron streams, the
final hidden vector retains the one-body form
$\bm h_i^L=\bm h^L(\bm r_i)$.

For determinant $m$, the element in electron row $i$ and orbital
column $j$ is
\begin{equation}
\phi_j^m(\bm r_i)
=\chi_j(\bm r_i)
 \bigl[\bm w_j^m\cdot\bm h_i^L(\bm r_i)\bigr],
\label{eq:complex-slater-orbitals}
\end{equation}
where $\bm w_j^m\in\mathbb C^{d_{\rm L}}$. The orbitals need not be orthogonal,
although linearly dependent orbitals make the determinant vanish. With $D$
determinants, the wavefunction is
\begin{equation}
\Psi(\bm R)
=\sum_{m=1}^{D}
  \det\!\left[\phi_j^m(\bm r_i)\right]_{i,j=1}^{N} .
\label{eq:complex-slater-readout}
\end{equation}
For $D>1$, the determinant values are summed directly without independent
mixing coefficients, as written in Equation~\eqref{eq:complex-slater-readout}.
Exchanging two electrons exchanges two rows of every Slater matrix and
changes the sign of the wavefunction. Translating electron $i$ by $\bm L_a$
leaves the periodic coefficient unchanged. It multiplies every element of row
$i$ by the common phase $e^{\ii\Lambda_a(\bm r_i)}$. The determinant therefore
satisfies the many-body magnetic boundary condition in
Equation~\eqref{eq:magnetic-boundary}. For $D=1$, Complex SlaterNet is the neural analogue of an unrestricted
single-determinant ansatz. Increasing $D$ gives a finite multideterminant
expansion without changing the one-body dependence in Equation~\eqref{eq:complex-slater-orbitals}. Complex SlaterNet therefore
provides the attention-free reference for introducing
configuration-dependent orbitals.

\subsection{Complex Psiformer wavefunction}
\label{sec:ansatz}

Self-attention is central to Transformer architectures
\cite{vaswani2017attention}. Complex-valued variants use real--imaginary
channels, complex projections, or Hermitian-magnitude similarities for
sequence and signal data
\cite{yang2020complex,eilers2023building,peng2024signal}. We adapt this
Hermitian-magnitude score and complex-value aggregation pattern to
permutation-equivariant electron streams and call the resulting block HMCVA.
This existing attention pattern couples the periodic complex coefficients
before the determinant, while the fixed magnetic sections enforce the torus
transition law.

Complex SlaterNet constructs each orbital value from one electron
stream. Following Ref.~\cite{geier2025attention}, attention extends this
one-body orbital according to
$\phi_j^m(\bm r_i)\rightarrow
\phi_j^m(\bm r_i;\{\bm r_{/i}\})$, where
$\{\bm r_{/i}\}=\{\bm r_k:k\ne i\}$ denotes the positions of all other
electrons. The coordinate before the semicolon identifies the electron
row, while the coordinates after it describe the many-electron environment.
HMCVA makes $\bm h_i^L$ a permutation-equivariant function of the full
configuration. The resulting orbitals carry many-electron dependence, as in neural
backflow and self-attention wavefunctions
\cite{pfau2020ferminet,vonglehn2023self,geier2025attention}.

At each layer, HMCVA exchanges information between electron streams
before the residual feed-forward update.
\begin{equation*}
\{\bm h_i^l\}_{i=1}^{N}
\xrightarrow{\mathrm{HMCVA}}
\{\bm f_i^{l+1}\}_{i=1}^{N}
\xrightarrow{\mathrm{feed\text{-}forward}}
\{\bm h_i^{l+1}\}_{i=1}^{N} .
\end{equation*}
We write the attention residual as
$\bm f_i^{l+1}=\bm h_i^l+\bm a_i^l$, where $\bm a_i^l$ is the multi-head update
defined below. The feed-forward layer then gives
$\bm h_i^{l+1}=\bm f_i^{l+1}+
\sigma^{l+1}(\mathcal F^{l+1}(\bm f_i^{l+1}))$.

At layer $l$, the index $h=1,\ldots,N_{\rm heads}$ labels an attention
head. Independent bias-free real-linear maps produce a query, a key, and a
value for each electron stream.
\begin{equation}
\begin{array}{rcl}
\bm q_i^{lh}&=&W_q^{lh}\bm h_i^l+V_q^{lh}\overline{\bm h_i^l},\\
\bm k_i^{lh}&=&W_k^{lh}\bm h_i^l+V_k^{lh}\overline{\bm h_i^l},\\
\bm v_i^{lh}&=&W_v^{lh}\bm h_i^l+V_v^{lh}\overline{\bm h_i^l}.
\end{array}
\label{eq:hmcva-qkv}
\end{equation}
The query $\bm q_i^{lh}\in\mathbb C^{d_q}$ specifies the learned
features sought by the receiving stream $i$. The key
$\bm k_j^{lh}\in\mathbb C^{d_k}$ describes how source stream $j$ matches that
query. The value $\bm v_j^{lh}\in\mathbb C^{d_v}$ carries the complex message
from the source. The trainable matrices $W_q^{lh},V_q^{lh}$,
$W_k^{lh},V_k^{lh}$, and $W_v^{lh},V_v^{lh}$ have shapes
$d_q\times d_{\rm L}$, $d_k\times d_{\rm L}$, and
$d_v\times d_{\rm L}$, respectively, over $\mathbb C$. The query and key share
the width $d_q=d_k$, while $d_v$ is independent.

HMCVA compares receiver $i$ with source $j$ through the magnitude of
their Hermitian dot product, scaled by the query width.
\begin{equation}
s_{ij}^{lh}
=\frac{|\overline{\bm q_i^{lh}}\cdot\bm k_j^{lh}|}{\sqrt{d_q}}.
\label{eq:hmcva-score}
\end{equation}
This score measures compatibility in the learned complex query--key
space. The absolute value removes the phase of the query--key overlap from the
real attention weight. Since the query and key maps are independent,
$s_{ij}^{lh}$ need not equal $s_{ji}^{lh}$. A larger $s_{ij}^{lh}$ gives source
$j$ more influence on the correlated representation of electron $i$, relative
to the other sources. This internal relevance measure has no direct interpretation as a
physical orbital overlap or interaction energy.

A softmax over all source electrons converts the scores into attention
weights. These weights then combine the complex values.
\begin{equation}
\alpha_{ij}^{lh}
=\frac{e^{s_{ij}^{lh}}}{\sum_{n=1}^{N}e^{s_{in}^{lh}}},\qquad
\operatorname{HEAD}_i^{lh}=\sum_{j=1}^{N}\alpha_{ij}^{lh}\bm v_j^{lh} .
\label{eq:hmcva-aggregation}
\end{equation}
The weights $\alpha_{ij}^{lh}$ are real, non-negative, and normalized
over $j$. They include the self-pair $j=i$. The weighted sum
$\operatorname{HEAD}_i^{lh}\in\mathbb C^{d_v}$ is the output of head $h$ for
electron $i$. Because the weights are real, the output retains the amplitude
and phase information carried by the complex values.
Appendix~\ref{app:structural-properties} discusses the symmetry consequences
of this score.

The $N_{\rm heads}$ outputs are concatenated in
$\mathbb C^{N_{\rm heads}d_v}$ and mapped back to
$\mathbb C^{d_{\rm L}}$ by a bias-free real-linear output map with
$W_{\rm o}^l,V_{\rm o}^l\in
\mathbb C^{d_{\rm L}\times(N_{\rm heads}d_v)}$. The projected output defines
the attention update $\bm a_i^l$ used above. All maps are shared across
electrons, and the sums over source streams do not depend on their ordering.
HMCVA is therefore permutation equivariant. After $L$ blocks,
$\bm h_i^L(\bm R)$ remains associated with electron $i$ while depending on the
full configuration.

Attention changes the periodic neural coefficient from a one-body
function of $\bm r_i$ to a permutation-equivariant function of the full
configuration, while preserving the magnetic orbital structure. Reusing the
same fixed $\chi_j$, the Complex Psiformer orbital is
\begin{equation}
\phi_j^m(\bm r_i;\{\bm r_{/i}\})
=\chi_j(\bm r_i)
  \bigl[\bm w_j^m\cdot\bm h_i^L(\bm R)\bigr] .
\label{eq:magnetic-orbitals}
\end{equation}
Equation~\eqref{eq:magnetic-orbitals} differs from the Complex
SlaterNet orbital in Equation~\eqref{eq:complex-slater-orbitals} through its
second argument. The magnetic factor $\chi_j(\bm r_i)$ remains one-body, while
$\bm h_i^L(\bm R)$ allows the periodic coefficient to respond to every other
electron. Replacing each one-body orbital value in
Equation~\eqref{eq:complex-slater-readout} with this correlated orbital gives
the Complex Psiformer wavefunction. The shared Slater structure
therefore preserves fermionic antisymmetry and the magnetic boundary condition,
as shown in Appendix~\ref{app:structural-properties}.

\section{Variational Monte Carlo}
\label{sec:vmc}

\subsection{Variational energy and Monte Carlo sampling}
\label{sec:vmc-sampling}

We optimize the real parameters $\bm\theta\in\mathbb R^P$ of the
complex trial wavefunction $\Psi_{\bm\theta}(\bm R)$ by variational Monte
Carlo (VMC)~\cite{foulkes2001qmc,geier2025attention}.
During training, Metropolis--Hastings sampling targets the density
$\mathsf p_{\bm\theta}(\bm R)\propto|\Psi_{\bm\theta}(\bm R)|^2$.
At configurations where $\Psi_{\bm\theta}(\bm R)\ne0$, the local energy is
\[
E_{\mathrm{loc},\bm\theta}(\bm R)
=\Psi_{\bm\theta}^{-1}(\bm R)
[\hat H\Psi_{\bm\theta}](\bm R).
\]
Its expectation gives the variational energy,
\begin{equation}
E_{\bm\theta}
=\mathbb E_{\bm R\sim\mathsf p_{\bm\theta}}
\left[E_{\mathrm{loc},\bm\theta}(\bm R)\right].
\label{eq:local-energy}
\end{equation}
For a trial state in the operator domain of the self-adjoint
Hamiltonian $\hat H$, this expectation is real.
It bounds the ground-state energy from above, although individual
local-energy samples may be complex.

Locally away from nodes, the wavefunction can be written as
$\Psi_{\bm\theta}(\bm R)=e^{u_{\bm\theta}(\bm R)}
e^{\ii\phi_{\bm\theta}(\bm R)}$, where
$u_{\bm\theta}(\bm R)=\log|\Psi_{\bm\theta}(\bm R)|$ is the log amplitude and
$\phi_{\bm\theta}(\bm R)$ is the local phase.
Sampling depends only on $u_{\bm\theta}$, while the kinetic local
energy requires coordinate derivatives of both $u_{\bm\theta}$ and
$\phi_{\bm\theta}$. Although exact nodes carry zero sampling probability,
nearby configurations can produce large local-energy fluctuations.
The independent Psiformer comparisons use the magnetic quadratic form under
the conditions and sampling protocol specified in Appendix~\ref{app:complex-vmc}.

\subsection{Stochastic reconfiguration and MinSR}
\label{sec:minsr}

All four models use stochastic reconfiguration
(SR)~\cite{sorella1998stochastic} through the same minimum-step SR (MinSR)
implementation~\cite{chen2024minsr,rende2024sridentity}.

For a batch $\{\bm R_b\}_{b=1}^{M}$, we collect the log-amplitude and
phase derivatives in $J^u,J^\phi\in\mathbb R^{M\times P}$, with
$J^u_{b\mu}=\partial_{\theta_\mu}u_{\bm\theta}(\bm R_b)$ and
$J^\phi_{b\mu}=\partial_{\theta_\mu}\phi_{\bm\theta}(\bm R_b)$. Away from
nodes and phase-branch discontinuities, they satisfy
\begin{equation}
\Psi_{\bm\theta}^{-1}(\bm R_b)
\partial_{\theta_\mu}\Psi_{\bm\theta}(\bm R_b)
=J^u_{b\mu}+\ii J^\phi_{b\mu}.
\label{eq:parameter-jacobian}
\end{equation}
For a differentiable trial family in the operator domain, the
energy gradient is a covariance of these responses with the local energy.
This expression assumes a parameter-independent Hamiltonian, as derived in
Appendix~\ref{app:complex-vmc}. The centered amplitude and phase responses
couple to the real and imaginary local-energy residuals, respectively.

For a batch of $M$ configurations, we stack the centered amplitude and
phase Jacobians into one real matrix,
\begin{equation}
X=\frac{1}{\sqrt M}
\begin{pmatrix}
J^u-\overline J^u\\
J^\phi-\overline J^\phi
\end{pmatrix}
\in\mathbb R^{2M\times P}.
\label{eq:full-complex-jacobian}
\end{equation}
Here the bars denote column-wise batch means.
The vector $\bm y\in\mathbb R^{2M}$ stacks the centered real and imaginary
local-energy residuals with the same $M^{-1/2}$ scaling.
The MinSR parameter update is
\begin{equation}
\Delta\bm\theta
=-2\eta X^{\mathsf T}
(XX^{\mathsf T}+\lambda I)^{-1}\bm y,
\label{eq:minsr-update}
\end{equation}
where $\eta$ is the learning rate, $\lambda$ is the damping, and $I$ is
the identity in sample space.
The linear solve has dimension $2M$ instead of $P$.
In training, energy clipping and an SR-metric norm constraint modify the
update, as detailed in Appendix~\ref{app:complex-vmc}.
We use Kronecker-factored approximate-curvature (KFAC) optimization only
for the auxiliary comparison in that appendix~\cite{martens2015kfac,pfau2020ferminet}.
Network and optimization settings are listed in
Tables~\ref{tab:slater_hyperparameters}, \ref{tab:psiformer_hyperparameters},
and \ref{tab:shared_hyperparameters}.


\section{Main results}
\label{sec:main-results}

We study $N=12$ electrons in a fixed spin-polarized sector, including the
periodic Ewald interaction and its Madelung term. The $25$-cell system uses a
$5\times5$ moir\'e supercell, while the $18$-cell system uses a $3\times6$
supercell. Their electron fillings per moir\'e cell are $12/25$ and
$2/3$, respectively. Both benchmarks have $n_{\phi}=2$ and use the
cell-centered symmetric gauge and magnetic boundary conditions defined in
Section~\ref{sec:system}. They differ in both electron filling and flux density,
so they do not form a fixed-density finite-size sequence. The zero-field member
of the flux scan uses ordinary periodic boundary conditions.

We use the $\mathrm{WSe}_2/\mathrm{WS}_2$ continuum-model parameters
of Ref.~\cite{geier2025attention}. They are $m^*/m_e=0.35$,
$V_0=15\,\mathrm{meV}$, $\varphi=\pi/4$,
$a_{\mathrm M}=8.031\,\mathrm{nm}$, and $\epsilon=5.0$. Length and energy are
measured in the effective Bohr radius and Hartree, respectively, with
$a_B^*=\frac{\epsilon m_e}{m^*}a_B$ and
$\mathrm{Ha}^*=\frac{m^*}{m_e\epsilon^2}\mathrm{Ha}$.

For $a_{\mathrm M}=8.031\,\mathrm{nm}$, the moir\'e unit-cell area is
$55.9\,\mathrm{nm}^2$. The $25$- and $18$-cell benchmarks contain $0.080$ and
$0.111$ flux quanta per moir\'e cell. These values correspond to laboratory
fields of approximately $5.9$ and $8.2\,\mathrm T$, respectively. The
$n_\phi=0,\ldots,4$ scan of the $18$-cell system spans zero field and finite
fields from approximately $4.1$ to $16.5\,\mathrm T$. Comparable fields have been
used to study competing charge-ordered and Hofstadter states in twisted
$\mathrm{WSe}_2/\mathrm{MoSe}_2$ \cite{kometter2023hofstadter}. The material and
filling differ, so this comparison provides physical context rather than a
direct experimental benchmark.

Both finite-field benchmarks have $N/n_\phi=6$, the number of electrons per
magnetic flux quantum. This continuum Landau-level filling differs from the
electron filling per moir\'e cell given above. The cyclotron scales are about
$2.0\,\mathrm{meV}$ at $5.9\,\mathrm T$ and $2.7\,\mathrm{meV}$ at
$8.2\,\mathrm T$, both below $V_0=15\,\mathrm{meV}$. The moir\'e potential thus
remains important alongside magnetic quantization, placing these calculations
in a moir\'e--Hofstadter regime.

\subsection{Variational energy benchmark}
\label{sec:variational-benchmark}

We compare the variational energies of Real and Complex Psiformers
with each other and with single-determinant neural Hartree--Fock (HF) baselines. The Real and Complex $D=1$ SlaterNets
use configuration-independent orbitals and provide the neural HF baselines.
Their finite neural parameterization and optimization need not attain the
HF minimum. Energy lowering relative to these baselines assesses the benefit
of the more flexible many-electron ansatz, whereas the direct Psiformer
comparison assesses the two complete architectures.

We compare the $D=4$ Real and Complex Psiformers at the same network depth,
hidden real dimension, determinant count, MinSR schedule, and 10,000-step
optimization budget. Their activations and attention rules remain model
specific, and Complex Psiformer has slightly fewer trainable parameters.
Both architectures produce complex wavefunctions. Their structural
differences and training settings are detailed in
Appendices~\ref{app:variational-review} and~\ref{app:hyperparameters}.

\begin{figure}[h]
\centering
\includegraphics[width=\linewidth]{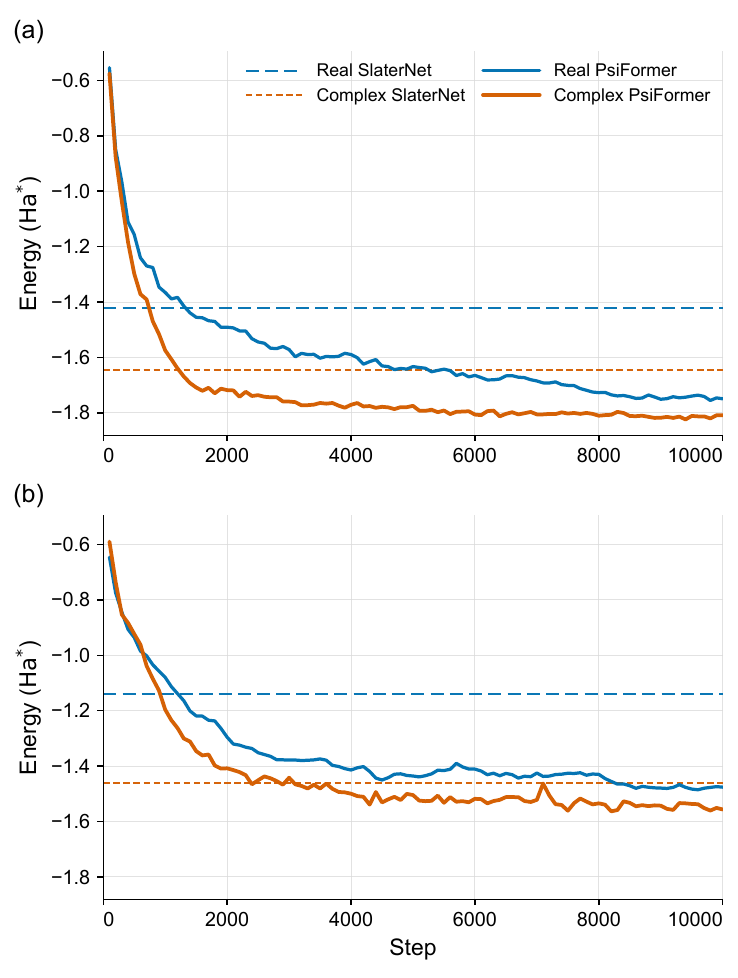}
\caption{Variational-energy optimization over the same number of steps.
Panels (a) and (b) show the $18$- and $25$-cell systems,
respectively, both with $N=12$ and $n_\phi=2$. Solid blue and orange curves show
the Real and Complex Psiformer training diagnostics.
Dashed blue and orange lines indicate the corresponding Real and Complex
single-determinant SlaterNet neural HF baselines. Table~\ref{tab:hf-benchmark}
reports the conventional local-energy comparison. The independent Psiformer
estimates discussed in the text use the magnetic quadratic-form protocol in
Appendix~\ref{app:complex-vmc}.}
\label{fig:energy-convergence}
\end{figure}

Figure~\ref{fig:energy-convergence} shows the training diagnostics.
Table~\ref{tab:hf-benchmark} reports conventional local-energy
estimates as total energies in $\mathrm{Ha}^*$.
We express the energy reductions below in meV per electron. Complex SlaterNet has
the lower mean estimate of the two neural HF baselines in both supercells.
Using this lower HF estimate as the common reference, the energy
reductions are $3.079$ and $0.508\,\mathrm{meV}$ per electron for
Real Psiformer in the $18$- and $25$-cell systems, respectively, and $5.016$
and $3.524\,\mathrm{meV}$ per electron for Complex Psiformer. The additional
variational flexibility comes from configuration-dependent orbitals and the
four-determinant expansion, as discussed in
Appendix~\ref{app:correlation-mechanisms}.

\begin{table}[h]
\centering
\caption{Conventional local-energy comparison with neural HF
baselines, in total $\mathrm{Ha}^*$, for $N=12$ and $n_\phi=2$ at 10,000
optimization steps. The SlaterNets use $D=1$ and the Psiformers $D=4$.
The hidden width $d_{\rm L}$ is stated over the native number field.
The quoted batch uncertainties are not autocorrelation-corrected errors
of independent energy estimates. Their interpretation is discussed in
Appendix~\ref{app:complex-vmc}.}
\label{tab:hf-benchmark}
\footnotesize
\setlength{\tabcolsep}{3pt}
\renewcommand{\arraystretch}{1.08}
\begin{tabular}{@{}lccc@{}}
\hline\hline
Model & $d_{\rm L}$ & $18$-cell & $25$-cell \\
\hline
Real SlaterNet
& $64\,(\mathbb R)$
& $-1.423 \pm 0.023$
& $-1.140 \pm 0.021$ \\
Complex SlaterNet
& $32\,(\mathbb C)$
& $-1.646 \pm 0.033$
& $-1.460 \pm 0.130$ \\
Real Psiformer
& $64\,(\mathbb R)$
& $-1.743 \pm 0.014$
& $-1.476 \pm 0.014$ \\
Complex Psiformer
& $32\,(\mathbb C)$
& $-1.804 \pm 0.009$
& $-1.571 \pm 0.019$ \\
\hline\hline
\end{tabular}
\end{table}

\begin{figure*}[t]
\centering
\includegraphics[width=0.92\textwidth]{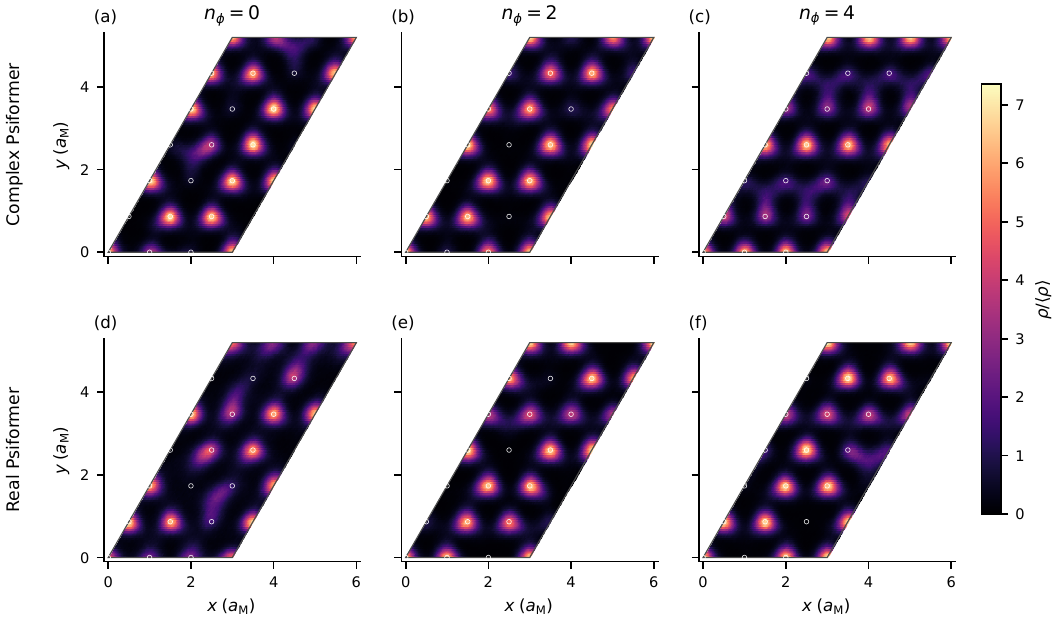}
\caption{Flux-resolved density of the optimized Psiformer states. Rows show Complex and Real Psiformer, with columns corresponding to $n_\phi=0$, $2$, and $4$ for the $18$-cell, $N=12$ system. The density is normalized by its spatial mean $\langle\rho\rangle=N/\Omega$. Each $96\times96$ density grid is mapped to $\bm r=u_1\bm L_1+u_2\bm L_2$, with $0\leq u_1,u_2<1$. The Cartesian axes $x$ and $y$ are in units of $a_{\mathrm M}$ and have equal scales, with the origin at the lower-left cell corner. Open circles mark minima of the moir\'e potential. Each panel uses 393,216 configurations sampled from a frozen state trained with seed $1$, without translation or rotation alignment or smoothing. The common color scale covers the full density range. At $n_\phi=2$, panel (b) exhibits a clearer honeycomb-type arrangement than panel (e).}
\label{fig:density-comparison}
\end{figure*}

To assess sensitivity to initialization, we trained each Psiformer with five
seeds in the $25$-cell system under the same 10,000-step budget. We then fixed
the trained parameters and evaluated each state with a separate Monte Carlo
calculation, following the common protocol in Appendix~\ref{app:complex-vmc}.
Across all five attempts, the mean energies are $-48.827$ and
$-47.369\,\mathrm{meV}$ per electron for Complex and Real Psiformer,
respectively. Complex Psiformer is lower by $1.458\,\mathrm{meV}$ per electron
on average and in four of the five comparisons by seed index. The sample standard
deviations across training seeds are $0.415$ and $1.000\,\mathrm{meV}$ per
electron, respectively. 

As a controlled small-system comparison, we also optimized the $18$-cell,
$N=2$, $n_\phi=2$ problem and constructed a finite configuration-interaction
(CI) reference using 40 active orbitals and all 780 determinants. The CI energy
is $-0.182\,\mathrm{Ha}^*$ in total, or
$-34.675\,\mathrm{meV}$ per electron. The frozen Complex and Real endpoints
lie $5.708$ and $22.599\,\mathrm{meV}$ per electron above that reference,
respectively. Thus the Complex endpoint is $16.891\,\mathrm{meV}$ per electron
below the Real endpoint for this fixed training pair. The energy comparison and its sensitivity to the evaluation sampling distribution are
shown in appendix Figure~\ref{fig:n2-benchmark}.
Appendix~\ref{app:ci-reference} gives the reference construction and
cutoff checks. The CI result is a controlled finite-basis
reference, not an exact continuum energy.

The energy gains apply to the complete Complex Psiformer at the
tested settings. Isolating the contributions of the radial activation,
attention rule, and residual stabilization requires component comparisons
with comparable tuning budgets.

\subsection{Spatial density profiles}

The one-body density $\rho(\bm r)$ describes the charge distribution
within each supercell. It satisfies
$\int_\Omega\rho(\bm r)\,d^2\bm r=N$, and its spatial mean is
$\langle\rho\rangle=N/\Omega$. We plot $\rho/\langle\rho\rangle$, so that
unity denotes a uniform state and deviations from unity measure spatial
charge modulation. Figure~\ref{fig:density-comparison} compares the unaligned
Real and Complex Psiformer profiles in the physical parallelogram geometry.
At the $18$-cell filling $N/(n_1n_2)=2/3$, scanning tunnelling microscopy in
$\mathrm{WSe}_2/\mathrm{WS}_2$ has resolved a honeycomb generalized Wigner
crystal with two of the three triangular-lattice sublattices occupied
\cite{li2021imagingwigner}. The same charge pattern appears in zero-field
continuum calculations with our material parameters and $\epsilon=5$
\cite{geier2025attention}. An independent neural-wavefunction study also
finds a $C_3$-symmetric crystal at this filling \cite{li2025wigner}.

At the main benchmark flux $n_\phi=2$, Complex Psiformer exhibits a clearer
honeycomb-type charge arrangement than Real Psiformer
[Figure~\ref{fig:density-comparison}(b,e)]. Integrating the density over the
nearest-well regions and averaging within the three sublattices gives
occupations $(0.953,0.157,0.890)$ for Complex and $(0.933,0.576,0.491)$ for
Real Psiformer, in the same sublattice convention. The two highly occupied
sublattices and one depleted sublattice of the Complex state are closer to
the ideal honeycomb motif $(1,0,1)$. The sublattice construction and the
site-resolved comparison are specified in Appendix~\ref{app:charge-order}. At this flux, the lower-energy Complex state is also closer to the
honeycomb pattern identified in the literature. This
comparison is qualitative because the cited studies do not provide a
ground-state density at the present field of approximately $8.2\,\mathrm T$.

In the Complex states, the honeycomb pattern is clearest at $n_\phi=2$ and
weakens at $n_\phi=4$, although local density maxima persist
[Figure~\ref{fig:density-comparison} and Appendix~\ref{app:charge-order}].
In the fixed spin-polarized model, the field acts through orbital
motion and may change which charge arrangement best balances kinetic,
moir\'e-potential, and Coulomb energies. This offers a possible explanation
for the reduced honeycomb contrast, although the separate energy
contributions have not been compared. Field-dependent competition between charge order
and Hofstadter states has also been observed in semiconductor moir\'e
lattices in other filling and band regimes
\cite{kometter2023hofstadter}. The field response depends on the
regime, as magnetic fields can also stabilize Wigner crystals at low
Landau-level filling \cite{tsui2024magneticwigner}. For these independently
optimized finite-cell states, a loss of honeycomb contrast does not by itself
establish crystal melting.

In the $25$-cell system, the density anisotropy varies substantially across
the five training seeds (Appendix~\ref{app:charge-order}). The seeds show no consistent distinction between stripe-like Real states
and two-dimensional Complex states. This system has filling $12/25$, and its cell
geometry is incommensurate with the $K$ and $M$ stars. Stripe order is itself
observed in the experimental half-filled crystal \cite{li2021imagingwigner},
so a stripe-like density alone does not indicate an unphysical state.

\subsection{Gauge-invariant current flow}

\begin{figure}[t]
\centering
\includegraphics[width=\columnwidth]{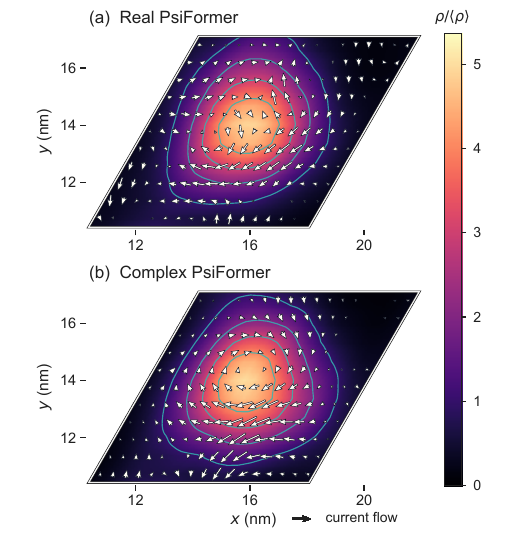}
\caption{Gauge-invariant current flow near a selected
density maximum. Panels (a) and (b) show the Real and
Complex Psiformers, respectively, for the $18$-cell, $N=12$, $n_\phi=2$
system after $10{,}000$ optimization steps. The background is the
relative one-body density $\rho/\langle\rho\rangle$, the cyan curves mark the
common contour levels $1$, $2$, $3$, and $4$, and the arrows show the
gauge-invariant mechanical particle current in
Eq.~\eqref{eq:gauge-invariant-current} rather than conventional charge current. Density color and current-arrow scales are shared between panels.
The same window in the $\bm L_1,\bm L_2$ supercell basis is shown in both
panels, with Cartesian coordinates given in nanometers. Each field is
evaluated on a $64\times64$ grid.
Periodic Gaussian smoothing with width $1.25$ grid pixels is applied only for
display.}
\label{fig:current-flow}
\end{figure}

The one-body density depends only on the wavefunction amplitude, whereas the
gauge-invariant mechanical one-body current probes its phase gradient. At fixed amplitude and vector potential, a change in the many-body phase can modify the current while leaving the density unchanged. We evaluate this current for the frozen $18$-cell, $N=12$, $n_\phi=2$ checkpoints. Using the local phase $\phi_{\bm\theta}(\bm R)$ defined above, the particle-current density in
effective atomic units is
\begin{equation}
\bm j(\bm r)=
\left\langle
\sum_{i=1}^{N}\delta_{\mathrm P}(\bm r-\bm r_i)
\left[\bm\nabla_i\phi_{\bm\theta}(\bm R)-\bm A(\bm r_i)\right]
\right\rangle_{|\Psi_{\bm\theta}|^2}.
\label{eq:gauge-invariant-current}
\end{equation}
Here $\delta_{\mathrm P}$ is the periodic delta function, and $\bm A$ is the cell-centered symmetric-gauge vector potential used in the Hamiltonian. Under a gauge transformation, the change in the phase gradient cancels that in the vector potential, leaving Equation~\eqref{eq:gauge-invariant-current} unchanged. Real Psiformer also produces a complex wavefunction from its real hidden features through a complex orbital readout. Both models include the prescribed magnetic section, so the current
probes the full optimized phase rather than the boundary factor alone. An exact
stationary state also satisfies $\bm\nabla\cdot\bm j=0$. This condition
permits local circulation without a change in the density.
Testing this continuity condition in a variational state requires
estimates over the full supercell with sampling uncertainties.

Figure~\ref{fig:current-flow} compares the two current fields after
$10{,}000$ optimization steps, using the same spatial grid, display scales,
and window near a selected density maximum. The two fields differ
in direction and magnitude near this maximum, with a clearer clockwise
circulation in the displayed Complex state.
These maps give a qualitative view of local circulation, which
depends on both the probability weights and phase gradients in
Equation~\eqref{eq:gauge-invariant-current}. Assessing current or phase
accuracy requires an independent reference. Transport and Hall responses
require separate response calculations.

\subsection{Flux-dependent density correlations}


We examine the flux dependence of density correlations in the $18$-cell
Complex Psiformer states optimized independently at $n_\phi=0,\ldots,4$.
The static structure factor resolves their equal-time density correlations
in momentum space,
\begin{equation}
S_{n_{\phi}}(\bm q)
=
\frac{1}{N}
\left\langle
\left|
\sum_{j=1}^{N}
e^{\ii\bm q\cdot\bm r_j}
\right|^2
\right\rangle_{n_{\phi}}.
\label{eq:structure-factor}
\end{equation}
The average is taken over the optimized state at each flux, with
$\bm q=m_1\bm G_1+m_2\bm G_2$, $m_1,m_2\in\mathbb Z$, chosen from the
supercell reciprocal lattice defined in Section~\ref{sec:system}.
Like the one-body density, $S(\bm q)$ depends
on $|\Psi|^2$, but it also resolves correlations between electron positions.
Using the independently optimized zero-field state as a reference, we define
\begin{equation}
\Delta S_{n_{\phi}}(\bm q)
=
S_{n_{\phi}}(\bm q)-S_{0}(\bm q).
\label{eq:structure-factor-difference}
\end{equation}

\begin{figure}[t]
\centering
\includegraphics[width=\linewidth]{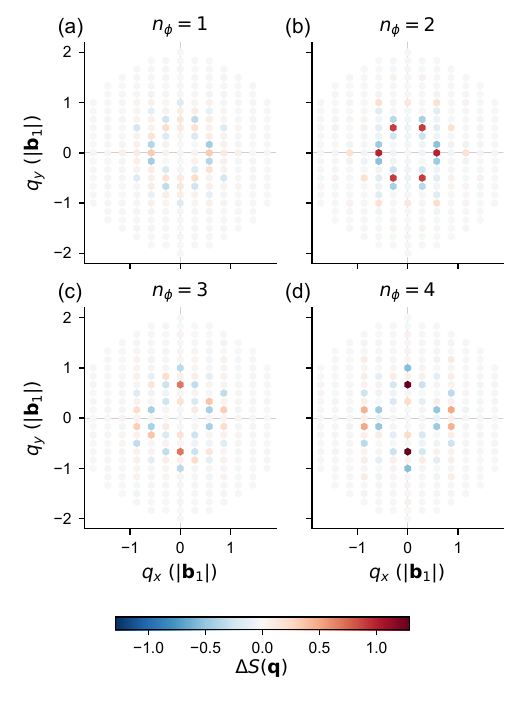}
\caption{Flux-dependent change in the static structure factor for
Complex Psiformer. Panels (a)--(d) show the $18$-cell, $N=12$ system at
$n_\phi=1$, $2$, $3$, and $4$, respectively, relative to the independently
optimized zero-field state. Positive and negative values indicate enhanced
and suppressed density correlations at the corresponding reciprocal-space
modes. The axis scale $|\bm b_1|$ is the magnitude of a primitive moir\'e
reciprocal vector, equal to $|\bm g_1|$ in the notation of the text.}
\label{fig:structure-factor-response}
\end{figure}

Figure~\ref{fig:structure-factor-response} shows changes concentrated at
selected reciprocal-space points, with both positive and negative
contributions relative to zero field. Different flux sectors emphasize
different modes, with the largest changes visible at $n_\phi=2$ and $4$. The magnitudes vary nonmonotonically with flux.

We separate this redistribution into changes in the average charge
arrangement and changes in density fluctuations. With
$\rho_{\bm q}=\sum_j e^{\ii\bm q\cdot\bm r_j}$ denoting the dimensionless
Fourier component of the microscopic number density, we write
\begin{equation}
S(\bm q)=\frac{|\langle\rho_{\bm q}\rangle|^2}{N}
+\frac{\langle|\rho_{\bm q}|^2\rangle
-|\langle\rho_{\bm q}\rangle|^2}{N},
\label{eq:structure-factor-decomposition}
\end{equation}
where the first term, denoted by $S_{\rm disc}(\bm q)$, is the disconnected
contribution from the mean density modulation, and the second term,
$S_{\rm conn}(\bm q)$, describes connected density fluctuations about that
mean. This density covariance can also be nonzero in a single
Slater determinant. Both contribute to the changes displayed in
Figure~\ref{fig:structure-factor-response}.

\begin{figure*}[t]
\centering
\includegraphics[width=0.94\textwidth]{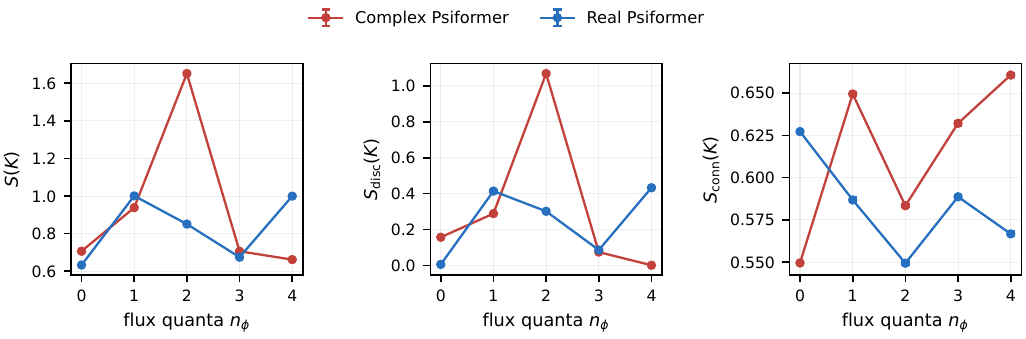}
\caption{Flux dependence of the $K$-star density correlations.
Panels show the total structure factor, its mean-density contribution, and
connected fluctuations, averaged over the six $K$-star vectors for the
$18$-cell Complex and Real Psiformer states. Error bars are time-block
standard errors from 393,216 configurations and are smaller than several
markers. }
\label{fig:structure-factor-components}
\end{figure*}

We quantify these contributions in an independent evaluation of both
Psiformers, averaging over the six moir\'e Brillouin-zone corners associated
with three-sublattice order. The resulting $S(K)$, $S_{\rm disc}(K)$, and
$S_{\rm conn}(K)$ are shown in Figure~\ref{fig:structure-factor-components}.
Appendix~\ref{app:charge-order} specifies the reciprocal vectors and
the averaging procedure. At
$n_\phi=2$, approximately $95.758\%$ of the Complex--Real difference in $S(K)$
comes from the mean-density contribution, consistent with the clearer
honeycomb arrangement in Figure~\ref{fig:density-comparison}(b). In the
Complex states, this contribution becomes small at $n_\phi=4$, while the
connected component remains finite and increases from its value at
$n_\phi=2$. The decrease in $S(K)$ is therefore dominated by the
loss of its mean-density contribution, despite the increase in connected
fluctuations. Averaging over translated charge arrangements can also weaken
the mean-density pattern while retaining crystalline correlations, as
discussed in Appendix~\ref{app:charge-order}.

Each point represents one independently optimized training
endpoint. The error bars in Figure~\ref{fig:structure-factor-components} describe sampling
of these fixed states and do not measure variation across training runs.
The fixed magnetic-factor assignment also changes with flux
(Appendix~\ref{app:magnetic-sections}). Establishing the ground-state field
response requires checks of both initialization and this factor choice.
The finite-cell comparison does not determine a thermodynamic phase boundary.

\section{Conclusion}
\label{sec:conclusion}

Complex Psiformer combines complex hidden features and HMCVA with
magnetic orbital sections for interacting continuum electrons.
Both Psiformers lower the energies relative to their neural HF baselines.
After the same number of optimization steps, Complex Psiformer reaches
lower energies than Real Psiformer in both finite-supercell benchmarks.
Its mean advantage in the $25$-cell system is $1.46\,\mathrm{meV}$ per
electron over all five attempted training seeds.
The separately trained $N=2$ Complex state also has a smaller energy gap
to the finite CI reference than its Real counterpart.

At the $18$-cell filling of $2/3$ and $n_\phi=2$, the lower-energy
Complex state has a clearer honeycomb-type density pattern, resembling the
charge motif reported at different fields. The larger $S(K)$ relative to
Real Psiformer mainly reflects the mean-density contribution.
Across the independently optimized Complex states, density correlations
vary nonmonotonically, with weaker honeycomb mean-density modulation at
$n_\phi=4$ while connected fluctuations persist. These observations from a
single finite cell do not establish crystal melting. The current maps provide
a qualitative comparison of local phase-dependent circulation.

The energy gains apply to the complete architecture under the
tested optimization settings. Component comparisons with matched tuning
and computation budgets are needed to identify their origin. Energy
decomposition and reference densities or currents would test the physical
consequences of the improvement. Alternative magnetic sections and repeated
optimizations would help assess the flux trends. Efficient derivative
evaluation and transferable training may extend these tests to larger
systems~\cite{li2024forwardlaplacian,nazaryan2026qernel,gaggioli2026largescale}.

\emph{Acknowledgments}. The authors sincerely thank Professor Hongjun Gao
for his guidance on the relevant literature and Professor Xi Dai for
insightful discussions. This work was partially supported by the National Natural Science Foundation of China (Grant No.92576114, 12447107), the Guangdong Provincial Quantum Science Strategic Initiative (Grant No.~GDZX2403008, GDZX2503001), and the Guangdong Provincial Key Lab of Integrated Communication, Sensing and Computation for Ubiquitous Internet of Things (Grant No.~2023B1212010007).

\emph{Statement on AI use.} OpenAI ChatGPT and Codex assisted with language
editing and the development and management of research code and experiments. The
authors supplied the problem formulation, source material, and revision
criteria. They independently
verified all mathematical statements, references, and final wording and
take full responsibility for the manuscript.

\emph{Data and code availability.} The model implementation is maintained at
\url{https://github.com/QuAIR/ComplexPsiformer} and is available from the
authors upon reasonable request.
Reference configuration files, individual-seed energies, sampling
diagnostics, and numerical-check files accompany the manuscript source.
Their contents, code versions, and software requirements are indexed in
\texttt{data/README.md}.
Training checkpoints and raw sampling trajectories are not included in
this source package.

\begin{appendix}

\section{From Slater determinants to neural attention wavefunctions}
\label{app:variational-review}

\subsection{Single-determinant reference and correlation energy}

For $N$ spin-polarized electrons, we suppress the common spin factor.
The Hartree--Fock (HF) reference is then a single determinant of orthonormal
spatial orbitals that satisfy the prescribed magnetic boundary conditions:
\begin{equation} 
\Psi_{\mathrm{HF}}(\bm R) =\frac{1}{\sqrt{N!}} \det\!\left[\phi_j(\bm r_i)\right]_{i,j=1}^{N}. 
\label{eq:hf-reference} 
\end{equation} 
Here $i$ labels an electron row, and $j$ labels an orbital column.
Exchanging two electron coordinates exchanges two rows and reverses the sign
of the wavefunction. Each matrix element depends only on $\bm r_i$, so
different electron coordinates couple only through antisymmetrization
\cite{slater1929complex,fock1930approximation}. This independent-orbital form
restricts the accessible many-body sign and phase structure. For real states,
it also restricts the nodal surface \cite{foulkes2001qmc}. Let $\mathcal S_1$
denote the nonzero single-determinant states in the same particle-number,
spin-polarization, and magnetic-boundary sector. If $E_0$ is the exact
ground-state energy in this sector, the HF and correlation energies are
\begin{equation} 
E_{\mathrm{HF}} =\min_{\Psi\in\mathcal S_1} \frac{\langle\Psi|\hat H|\Psi\rangle} {\langle\Psi|\Psi\rangle}, \qquad E_{\mathrm{corr}}=E_0-E_{\mathrm{HF}}\leq0. \label{eq:correlation-energy} \end{equation} 
By the variational principle, the exact correlation energy is non-positive
under the convention in Equation~\eqref{eq:correlation-energy} \cite{foulkes2001qmc,geier2025attention}. Energy
differences between finite variational models should therefore not be identified
with $E_{\mathrm{corr}}$. We compare variational energies directly for
models optimized with the same Hamiltonian and boundary conditions.
The $D=1$ SlaterNet with configuration-independent orbitals has the
single-determinant form of Equation~\eqref{eq:hf-reference}. Its restricted
neural parameterization and finite optimization need not attain
$E_{\mathrm{HF}}$. We therefore call the optimized Real and Complex $D=1$
SlaterNets neural HF baselines. For each system, the energy
reductions reported in Section~\ref{sec:variational-benchmark} use the lower
of the two neural HF estimates in Table~\ref{tab:hf-benchmark} as a common
reference. A multideterminant SlaterNet lies beyond $\mathcal S_1$.
A Psiformer generally lies beyond the HF class even with one determinant,
because its orbitals depend on the full electronic configuration.

\subsection{Correlation mechanisms and neural extensions}
\label{app:correlation-mechanisms}

Correlated fermionic ansatzes extend the independent-orbital determinant
by introducing many-electron dependence into selected parts of the
wavefunction. One representative form combines a Jastrow factor with a
determinant expansion:
\begin{equation*} 
\Psi(\bm R) =e^{J(\bm R)}\sum_{k=1}^{K} \det\!\left[ \phi_j^{(k)}\!\left(\bm q_i(\bm R)\right) \right]_{i,j=1}^{N}. 
\end{equation*} 
Here $J(\bm R)$ is exchange symmetric, and the backflow coordinate
$\bm q_i(\bm R)$ is permutation equivariant. Fermionic antisymmetry therefore
remains enforced by the determinants. For a real Jastrow factor,
$e^{J(\bm R)}>0$ changes the amplitude without altering the determinant zeros
or phase \cite{jastrow1955manybody,foulkes2001qmc}. Its short-range form can
also impose the Coulomb cusp condition \cite{kato1957eigenfunctions}. A
multideterminant expansion introduces interference between antisymmetric
components. Backflow instead makes each orbital coordinate depend on the full
configuration and can deform the determinant zero set
\cite{feynman1956spectrum,kwon1993backflow}. Continuous first-quantized neural
wavefunctions use closely related mechanisms. PauliNet combines multireference
orbitals with neural Jastrow and backflow factors. FermiNet constructs
determinants from configuration-dependent orbitals
\cite{hermann2020paulinet,pfau2020ferminet}.

Neural backflow provides an analogous construction in lattice
occupation space \cite{luo2019backflow}. Correlation may therefore enter
through symmetric many-body factors, determinant interference, or
configuration-dependent orbital values. The SlaterNet--Psiformer hierarchy
focuses on the last mechanism while retaining the same determinant-based
antisymmetric readout. SlaterNet applies a shared neural map independently to
each electron before the orbital readout \cite{geier2025attention}. Psiformer
adds self-attention between electron streams, so each orbital coefficient can
depend on the full configuration \cite{vaswani2017attention,vonglehn2023self}.
The determinant expansion remains unchanged, so the architectures differ
in how many-electron information reaches the orbital values before
antisymmetrization.

Fu showed that a continuous fermionic wavefunction on a compact
domain can be approximated by finite sums of antisymmetric basis functions
with symmetric coefficients~\cite{fu2026fermisets}.
Here the basis functions and their symmetric coefficients are both continuous.
The construction requires nontrivial antisymmetric basis functions that
remain nonzero away from particle coincidences.

Table~\ref{tab:related-magnetic-ansatzes} compares the geometry, boundary
conditions, and learned representation of related continuum wavefunctions.
Complex Psiformer extends the periodic moir\'e setting of
Ref.~\cite{geier2025attention} to a magnetic torus and uses complex electron
features and attention messages throughout the periodic coefficient network.

\begin{table*}[t]
\centering
\caption{Relation to attention-based continuum fermionic wavefunctions. The
entries describe the settings studied in each work.}
\label{tab:related-magnetic-ansatzes}
\footnotesize
\setlength{\tabcolsep}{4pt}
\renewcommand{\arraystretch}{1.12}
\begin{tabular}{@{}p{0.16\textwidth}p{0.17\textwidth}p{0.22\textwidth}
p{0.37\textwidth}@{}}
\hline\hline
Work & Geometry and field & Hidden representation & Boundary and physical setting \\
Geier \emph{et al.}~\cite{geier2025attention}
& Periodic supercell, zero field
& Real self-attention with complex orbital readout
& Periodic Fourier features for interacting electrons in a moir\'e potential \\
Teng \emph{et al.}~\cite{teng2025solving}
& Confined disk, finite field
& Real self-attention with complex generalized orbitals
& Gaussian orbital envelopes and a cusp Jastrow factor for an electron gas
with a charged-disk background \\
MagNet~\cite{abouelkomsan2026magnet}
& Torus, finite field
& Real self-attention with complex generalized orbitals
& Magnetic-translation boundary conditions for the homogeneous Coulomb
electron gas, including optional boundary twists \\
This work
& Periodic moir\'e supercell, finite field
& Complex hidden streams, HMCVA messages, and complex orbitals
& Fixed magnetic sections times learned periodic coefficients for a Coulomb
system in a triangular moir\'e potential \\
\hline\hline
\end{tabular}
\end{table*}

\subsection{Periodic and magnetic neural wavefunctions}

Periodic coordinate encoding, many-body correlation, and magnetic
boundary conditions serve distinct roles in an extended-system neural
wavefunction. For a supercell generated by $\bm L_1$ and $\bm L_2$, the
reciprocal vectors $\bm G_a$ satisfy
$\bm G_a\cdot\bm L_b=2\pi\delta_{ab}$. Periodic coordinates can therefore be
represented by Fourier features such as
\begin{equation*} 
\bm f^0(\bm r_i) = \left( e^{\ii\bm G_1\cdot\bm r_i}, e^{\ii\bm G_2\cdot\bm r_i} \right). 
\end{equation*} 
The equivalent sine--cosine representation is widely used in neural
wavefunctions for periodic solids
\cite{pescia2022neuralnetwork,cassella2023discovering}. These features encode
position on the finite torus without restricting the learned function to a
finite plane-wave expansion. Generalized orbitals, backflow, message passing,
pairing, and Bloch-based constructions can then add correlations to this
periodic representation
\cite{li2022deepsolid,wilson2023wapnet,pescia2024messagepassing,
luo2023pairing,luo2024neuralbloch}.

Within this hierarchy, SlaterNet and Psiformer differ in the periodic
neural coefficient. SlaterNet retains a one-body form, whereas self-attention
promotes the coefficient to a many-body representation. Shared attention maps
make $\bm h_i^L(\bm R)$ permutation equivariant. The coefficient associated
with electron $i$ can therefore depend on all other coordinates before
determinant antisymmetrization
\cite{vonglehn2023self,geier2025attention}. Because the network uses periodic
coordinate features, translating any electron by a supercell vector leaves the
learned coefficient unchanged. This property holds for both real and complex
hidden representations.

\begin{figure}[h]
    \centering
    \includegraphics[width=0.77\linewidth]{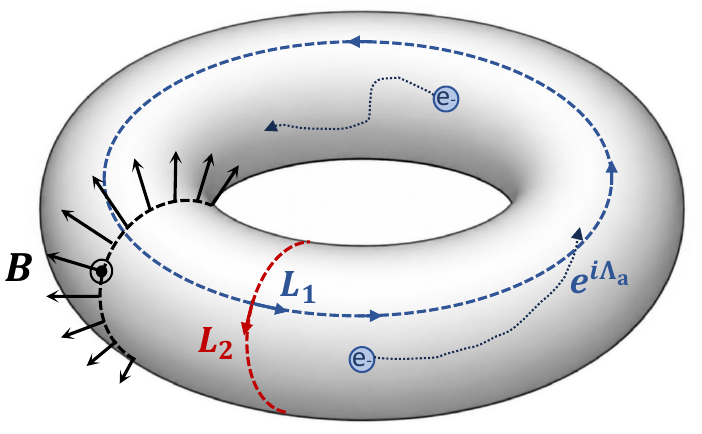}
    \caption{Torus representation of a magnetic supercell. In a periodic cell threaded by magnetic flux, a supercell translation changes the vector potential by a gauge transformation. The periodic neural coefficient is unchanged by the translation, while the full orbital is patched across opposite edges by the phase $e^{i\Lambda_a(\bm r)}$.}
    \label{fig:magnetic_gauge}
\end{figure}

At nonzero magnetic flux, as shown in Figure~\ref{fig:magnetic_gauge}, a uniform-field vector potential cannot be
globally periodic on a torus carrying nonzero flux. Under a supercell
translation, it instead returns to an equivalent gauge,
$\bm A(\bm r+\bm L_a)=\bm A(\bm r)+\bm\nabla\Lambda_a(\bm r)$. Gauge covariance
of the minimally coupled kinetic energy then requires a position-dependent
wavefunction phase rather than ordinary periodicity
\cite{onofri2001landau,haldane1985translational}. We impose this structure
through fixed magnetic sections $\chi_j$ satisfying
$\chi_j(\bm r_i+\bm L_a) = e^{\ii\Lambda_a(\bm r_i)}\chi_j(\bm r_i)$. The
learned coefficient carries the variational many-body dependence, while
$\chi_j$ supplies the prescribed magnetic transition law. At $n_\phi=0$, the
section is omitted and the periodic neural coefficient directly supplies the
orbital.

Previous neural wavefunctions have introduced magnetic structure through
geometry-dependent envelopes, complex-valued factors, winding-aware readouts,
or monopole harmonics
\cite{teng2025solving,perezfadon2025anyon,abouelkomsan2026magnet,
qian2025landau}. Winding-based constructions can also make the zeros of the
magnetic orbital factor depend on the electronic configuration
\cite{abouelkomsan2026magnet}. At finite field, our hierarchy keeps $\chi_j$
fixed across all models and places the trainable many-electron dependence in the
periodic coefficient. This coefficient may use real or complex features and
may be attention free or configuration dependent. The model-specific attention
rules and common MinSR settings are described below and in
Appendix~\ref{app:hyperparameters}.

\subsection{Representations of neural network ansatzes}

On a compact domain, a feed-forward neural network with a suitable
nonlinear activation can approximate any continuous real-valued function
arbitrarily well~\cite{hornik1989multilayer}.
Applying this result separately to the real and imaginary parts gives a
neural representation of continuous complex orbitals.

The models differ in the number field of the hidden electron
representation and in whether electron streams communicate before
the orbital readout. Real and Complex
SlaterNets process electrons independently, whereas Real and Complex
Psiformers communicate through attention. Within each architecture, the real
and complex variants differ in their hidden parameterization. The final
electron representation depends on the coordinates as
\begin{equation} 
\bm h_i^L= \begin{cases} 
\bm h^L(\bm r_i), & \text{SlaterNets},\\[3pt] 
\bm h_i^L(\bm R), & \text{Psiformers}. 
\end{cases} 
\label{eq:model-hierarchy-hidden} 
\end{equation} 
The attention-free models therefore retain one-body orbitals even when
several determinants are used. In Psiformer, attention makes the coefficient
associated with electron $i$ depend on the full configuration. The determinant
readout enforces fermionic antisymmetry in both cases. The real models use
$\bm h_i^L\in\mathbb R^{d_{\rm L}}$ and two real readouts to form a complex
orbital coefficient. The complex models instead use
$\bm h_i^L\in\mathbb C^{d_{\rm L}}$ and one complex readout. With the same
magnetic factor $\chi_j$, the orbital values are
\begin{equation*} 
\phi_j^m = \chi_j(\bm r_i) 
\begin{cases} 
\bm w_{j,\mathrm R}^{m}\cdot\bm h_i^L +\ii\bm w_{j,\mathrm I}^{m}\cdot\bm h_i^L, & \bm h_i^L\in\mathbb R^{d_{\rm L}},\\[4pt] 
\bm w_j^m\cdot\bm h_i^L, & \bm h_i^L\in\mathbb C^{d_{\rm L}}. 
\end{cases} 
\end{equation*} 
The argument of $\bm h_i^L$ follows
Equation~\eqref{eq:model-hierarchy-hidden}. Both model classes can therefore
produce complex orbital values. They differ in whether complex structure is
retained within the hidden representation and, for Psiformer, within messages
between electron streams \cite{geier2025attention}. One complex channel
contains two real degrees of freedom. We therefore compare a complex width
$d_{\rm L}$ with a real width $2d_{\rm L}$. This matching is consistent with
the real-affine complex layers used here. For the map in
Equation~\eqref{eq:free-real-linear}, with input and output widths
$d_{\mathrm{in}}$ and $d_{\mathrm{out}}$, the number of real parameters is
$4d_{\mathrm{out}}d_{\mathrm{in}} +2d_{\mathrm{out}}$. 
The map is equivalent to an unrestricted affine map from
$\mathbb R^{2d_{\mathrm{in}}}$ to $\mathbb R^{2d_{\mathrm{out}}}$,
with the same number of real parameters. The final
term is absent for a bias-free map. The Psiformer comparison therefore matches
the underlying hidden real dimensions without imposing equal total parameter
counts. Appendix~\ref{app:hyperparameters} reports the complete counts.

The radial activation and HMCVA can likewise be evaluated using
paired real and imaginary components without changing the represented
function. We checked an implementation using only real tensors throughout
the network, magnetic factors, and determinant sum. In 24 synthetic cases
with identical real parameter values, covering $N=2,12$, $n_\phi=0,2$,
and both branches of the residual limiter, the log-amplitude, unit phase,
coordinate gradients, and gradients with respect to all real parameters
agreed in double precision. The largest absolute difference in the parameter
gradients was $2.74\times10^{-12}$. Full coordinate Hessians also agreed in
four magnetic cases away from singular and switching points. These checks
verify agreement at the tested configurations.
The Real Psiformer baseline uses different nonlinearities and attention
rules, as described below.

With the layer and HEAD indices suppressed, the attention message path
in Real Psiformer is
\begin{equation*}
\alpha_{ij}^{\mathrm R}
=\softmax_j\!\left[(\bm q_i^{\mathrm R})^{\mathsf T}\bm k_j^{\mathrm R}\right],
\quad
\operatorname{HEAD}_i^{\mathrm R}
=\frac{1}{\sqrt{d_v}}\sum_j\alpha_{ij}^{\mathrm R}\bm v_j^{\mathrm R}.
\end{equation*}
Complex Psiformer instead uses
\begin{equation*}
\alpha_{ij}^{\mathrm C}
=\softmax_j\!\left[
\frac{|(\bm q_i^{\mathrm C})^\dagger\bm k_j^{\mathrm C}|}{\sqrt{d_q}}
\right],
\quad
\operatorname{HEAD}_i^{\mathrm C}
=\sum_j\alpha_{ij}^{\mathrm C}\bm v_j^{\mathrm C}.
\end{equation*}
Both rules produce real, non-negative attention weights. Real
Psiformer communicates real-valued features. HMCVA instead transports complex
values through the value and residual streams.
The Real Psiformer normalization follows
Ref.~\cite{geier2025attention}. It applies softmax to the unscaled real dot
product and multiplies the aggregated value by $d_v^{-1/2}$.
Scaling the score before softmax changes the relative attention weights.
Scaling the aggregated value leaves those weights unchanged and changes
the message amplitude. The HMCVA invariance discussed in
Appendix~\ref{app:structural-properties} applies only to the scalar query--key
compatibility. It does not imply gauge equivariance of the full network. At
finite field, all models use the same nontrainable magnetic sections. The
neural HF baselines use one determinant, while the Psiformers use four. The
Psiformer pair matches the hidden real dimension, depth, and determinant count.
The feature number field, activation, and attention rule remain model specific,
and all models use MinSR. These settings define a comparison of the complete architectures.

\section{Periodic Coulomb interaction}
\label{app:ewald}

The electrons are confined to two dimensions and interact through the
Coulomb potential $1/r$. In a periodic supercell, we evaluate this interaction
with an Ewald decomposition and a uniform neutralizing background
\cite{fraser1996finitesize,foulkes2001qmc,geier2025attention}.

Let $\Lambda=\{n_1\bm L_1+n_2\bm L_2:n_1,n_2\in\mathbb Z\}$ denote
the supercell lattice and
$\Lambda^\ast=\{m_1\bm G_1+m_2\bm G_2:m_1,m_2\in\mathbb Z\}$ its reciprocal
lattice. The Ewald splitting length $\eta_{\mathrm{Ew}}>0$ separates the real-
and reciprocal-space contributions. The resulting periodic pair interaction
is
\begin{equation*}
\begin{aligned}
v_{\mathrm{Ew}}(\bm r)
={}&
\sum_{\bm L\in\Lambda}
\frac{\operatorname{erfc}\!\left(
|\bm r+\bm L|/(2\eta_{\mathrm{Ew}})
\right)}
{|\bm r+\bm L|} \\
&+
\frac{2\pi}{\Omega}
\sum_{\substack{\bm G\in\Lambda^\ast\\ \bm G\neq\bm 0}}
\frac{\operatorname{erfc}(\eta_{\mathrm{Ew}}|\bm G|)}
{|\bm G|}
\cos(\bm G\cdot\bm r)
-
\frac{4\sqrt{\pi}\eta_{\mathrm{Ew}}}{\Omega}.
\end{aligned}
\end{equation*}
The neutralizing background removes the reciprocal zero mode and
produces the final constant term.

The regular part of the periodic interaction at the origin defines the
corresponding Madelung constant,
\begin{equation*}
\xi_{\mathrm M}
={}
\lim_{\bm r\to\bm 0}
\left[
v_{\mathrm{Ew}}(\bm r)-\frac{1}{|\bm r|}
\right].
\end{equation*}
The complete periodic Coulomb contribution to
Equation~\eqref{eq:hamiltonian} is
\begin{equation*}
V_{\mathrm{ee}}(\bm R)
=
\sum_{i<j}
v_{\mathrm{Ew}}(\bm r_i-\bm r_j)
+
\frac{N}{2}\xi_{\mathrm M}.
\end{equation*}

We checked the Ewald truncation on 256 fixed $N=12$ configurations
from a scrambled Sobol sequence for each supercell. Increasing both index
cutoffs from 5 to 16 gave maximum changes in $V_{\mathrm{ee}}/N$ of
$2.07\times10^{-4}$ and $2.10\times10^{-8}\,\mathrm{meV}$ in the
$18$- and $25$-cell geometries, respectively. The cutoff-12 results agreed
with the reference to double-precision roundoff, as did a 25\% variation
of the splitting length at cutoff 16. These maxima apply to the fixed
configuration set used for this truncation check.

\section{Construction of the magnetic sections}
\label{app:magnetic-sections}

The neural orbital coefficients introduced in the main text are
periodic under supercell translations. At nonzero magnetic flux, the complete
orbital must also acquire the position-dependent phase in
Equation~\eqref{eq:magnetic-boundary}. We impose this boundary law through fixed
functions $\chi_j$. This appendix constructs them for the oblique supercells
used in our calculations, following the torus Landau-level representation of
Ref.~\cite{onofri2001landau}.

Let $\bm L_1$ and $\bm L_2$ denote the oriented supercell translation
vectors, with area $\Omega=|\bm L_1\times\bm L_2|>0$. The uniform magnetic
field is perpendicular to the plane, $\bm B=B\hat{\bm z}$. In the
charge-weighted convention of the Hamiltonian, $B>0$ is the field strength and
satisfies $B\Omega=2\pi n_{\phi}$. Here
$n_{\phi}\in\mathbb N_{>0}$ is the number of flux quanta through the
supercell. We center the symmetric gauge at
$\bm l=(\bm L_1+\bm L_2)/2$.

We resolve the oblique cell into directions parallel and perpendicular
to $\bm L_1$. The associated length and unit vectors are
\begin{equation*}
L_{\parallel}=|\bm L_1|,
\qquad
\bm e_{\parallel}=\frac{\bm L_1}{L_{\parallel}},
\qquad
\bm e_{\perp}=\hat{\bm z}\times\bm e_{\parallel}.
\end{equation*}
The shear $s$ and perpendicular cell length $L_{\perp}$ are
\begin{equation*}
s=\bm L_2\cdot\bm e_{\parallel},
\qquad
L_{\perp}=\bm L_2\cdot\bm e_{\perp}>0.
\end{equation*}
These definitions give
$\bm L_1=L_{\parallel}\bm e_{\parallel}$,
$\bm L_2=s\bm e_{\parallel}+L_{\perp}\bm e_{\perp}$, and
$\Omega=L_{\parallel}L_{\perp}$.

For the $N$-electron determinant, we assign two nonnegative integer
labels to each orbital column $j=1,\ldots,N$:
\begin{equation*}
\mu_j=(j-1)\bmod n_{\phi},
\qquad
\nu_j=\left\lfloor\frac{j-1}{n_{\phi}}\right\rfloor.
\end{equation*}
The label $\mu_j\in\{0,\ldots,n_{\phi}-1\}$ selects a
magnetic-translation sector, while $\nu_j$ selects a transverse oscillator
channel. For each integer image index $n\in\mathbb Z$, the longitudinal wave
number is
\begin{equation*}
k_{jn}=\frac{2\pi\mu_j}{L_{\parallel}}+B L_{\perp}n.
\end{equation*}
Here $k_{j0}=2\pi\mu_j/L_{\parallel}$ denotes the $n=0$ member of
this sequence.

All models use the same column assignment at a given flux.
Changing $n_\phi$ in the flux scan also changes the assigned oscillator
channels. At zero flux the magnetic factors are omitted. The scan does not
assess sensitivity to alternative assignments.

In a uniform field, the corresponding Landau-gauge functions separate
into a plane wave along $\bm e_{\parallel}$ and a harmonic-oscillator profile
along $\bm e_{\perp}$ \cite{onofri2001landau}. The channel $\nu_j$ selects the
normalized transverse profile
\begin{equation*}
\varphi_{\nu}(\xi)
=
\frac{H_{\nu}(\xi)e^{-\xi^2/2}}
{\left(2^{\nu}\nu!\sqrt{\pi}\right)^{1/2}},
\qquad
\xi\in\mathbb R,
\end{equation*}
where $H_{\nu}$ is the physicists' Hermite polynomial of order
$\nu=0,1,\ldots$.

Let $\overline{\bm r}$ denote the representative of $\bm r$ in the
canonical supercell centered at $\bm l$. Its scalar coordinates relative to
the gauge center are
\begin{equation*}
x_{\parallel}=(\overline{\bm r}-\bm l)\cdot\bm e_{\parallel},
\qquad
x_{\perp}=(\overline{\bm r}-\bm l)\cdot\bm e_{\perp}.
\end{equation*}
Within this canonical cell, the $j$th magnetic section is
\begin{equation}
\begin{multlined}
\widetilde\chi_j(\overline{\bm r})
=
\frac{B^{1/4}}{\sqrt{L_{\parallel}}}
\exp\!\left(\frac{\ii Bx_{\parallel}x_{\perp}}{2}\right)
\\
{}\times
\sum_{n\in\mathbb Z}
\varphi_{\nu_j}\!\left[
\sqrt B\left(x_{\perp}+\frac{k_{jn}}{B}\right)
\right]
\\
{}\times
\exp(\ii x_{\parallel}k_{jn})
\exp\!\left[
\ii\left(
n k_{j0}s+\frac{1}{2}B L_{\perp}s n^2
\right)
\right].
\end{multlined}
\label{eq:canonical-magnetic-section}
\end{equation}
The Hermite function and $\exp(\ii x_{\parallel}k_{jn})$ form the
Landau-gauge orbital. The final exponential accounts for the shear of the
oblique cell. The factor containing $x_{\parallel}x_{\perp}$ converts the
orbital to the cell-centered symmetric gauge used in the Hamiltonian.

The boundary phases follow directly from this infinite sum.
Under translation by $\bm L_1$, the identity
$k_{jn}L_{\parallel}=2\pi(\mu_j+n_\phi n)$ leaves each plane-wave factor
unchanged, and the prefactor supplies $e^{\ii B L_{\parallel}x_{\perp}/2}$.
Translation by $\bm L_2$ sends $(x_{\parallel},x_{\perp})$ to
$(x_{\parallel}+s,x_{\perp}+L_{\perp})$.
Using $k_{j,n+1}=k_{jn}+B L_{\perp}$ to shift the image index restores
the oscillator argument. Combining the remaining phases gives
$e^{\ii B(sx_{\perp}-L_{\perp}x_{\parallel})/2}$.
These are the factors $e^{\ii\Lambda_a(\overline{\bm r})}$ in
Equation~\eqref{eq:magnetic-boundary}. Differentiating the identities gives
the corresponding matching of covariant derivatives.

In the numerical implementation, we truncate the image sum to $|n|\leq8$,
as listed in Table~\ref{tab:shared_hyperparameters}.
At finite cutoff, shifting the image index leaves endpoint terms.
The winding construction below enforces whole-cell translation phases,
while matching across the cell seams depends on the truncated tails.

Increasing the cutoff to 12 and 16 gave agreement to double-precision
roundoff for the fixed functions and their first and second coordinate
derivatives. We tested both supercell geometries with $N=12$ and
$n_\phi=1,\ldots,4$, together with the $N=2$, $n_\phi=2$ benchmark.
Each case used 384 points, including 128 near the cell seams.
We also evaluated the finite sum at 16 points on each boundary.
After applying the boundary phase, the values, covariant gradients, and
covariant Laplacians matched to double-precision roundoff.
These checks concern the fixed magnetic
functions and do not bound errors in the full neural wavefunction.

To extend the section beyond the canonical cell, write an arbitrary
position as $\bm r=\overline{\bm r}+w_1\bm L_1+w_2\bm L_2$,
where $w_1,w_2\in\mathbb Z$ count windings along the two supercell
directions. For $a=1,2$, the symmetric-gauge transition function is
\begin{equation*}
\Lambda_a(\bm r)
=
\frac{B}{2}\left[\bm L_a\times(\bm r-\bm l)\right]_z.
\end{equation*}
A path that first winds along $\bm L_1$ and then along $\bm L_2$ gives
\begin{equation*}
\chi_j(\bm r)
=
\exp\!\left[
\ii w_1\Lambda_1(\overline{\bm r})
+
\ii w_2\Lambda_2(\overline{\bm r}+w_1\bm L_1)
\right]
\widetilde\chi_j(\overline{\bm r}).
\end{equation*}
The two possible winding orders differ by
\begin{equation*}
\frac{
e^{\ii\Lambda_2(\bm r+\bm L_1)}e^{\ii\Lambda_1(\bm r)}
}{
e^{\ii\Lambda_1(\bm r+\bm L_2)}e^{\ii\Lambda_2(\bm r)}
}
=e^{-\ii B\Omega}
=1.
\end{equation*}
Flux quantization makes the extension independent of the winding order
and yields the required transition law
\begin{equation}
\chi_j(\bm r+\bm L_a)
=
e^{\ii\Lambda_a(\bm r)}\chi_j(\bm r),
\qquad
a=1,2.
\label{eq:section-transition}
\end{equation}

The functions $\chi_j$ remain fixed during optimization and are shared
by all four neural models. They supply the magnetic boundary law, while the
learned orbital coefficients retain the variational dependence on electron
coordinates. Translating electron $i$ by $\bm L_a$ multiplies every element in
determinant row $i$ by the same factor
$e^{\ii\Lambda_a(\bm r_i)}$. The determinant therefore satisfies the
many-body magnetic boundary condition \cite{haldane1985translational}.
Although the construction uses Landau-level functions, the unrestricted
periodic neural coefficient does not project the variational orbital onto a
fixed Landau-level subspace.

For the frozen $25$-cell, $N=12$, $n_\phi=2$ seed-$1$ Psiformers,
we checked both primitive seams using two background configurations, moving
each electron in turn. The wavefunction and its covariant directional
derivatives were compared on opposite sides at fractional displacements
from $10^{-2}$ to $10^{-8}$ along the corresponding primitive vector.
At the smallest displacement, the largest value difference divided by
the larger wavefunction amplitude in each pair was $3.13\times10^{-6}$.
For both models, the maximum value mismatch decreased as the displacement
was reduced.
These finite probes assess local matching at the tested configurations.

\section{Structural properties of Complex Psiformer}
\label{app:structural-properties}

Multiplying either the query or the key by a unit complex phase leaves
the Hermitian-magnitude score in Equation~\eqref{eq:hmcva-score} unchanged. The
overlap phase therefore does not enter the real attention weight. Changes in
the query or key magnitudes do change the score. Phase information remains in
the complex values and hidden representations. This score invariance does not imply gauge equivariance of the full
network, whose projections are general real-linear maps.

We first verify exchange symmetry. Let $\Pi$ be an $N\times N$
permutation matrix that reorders the electron coordinates in
$\bm R=(\bm r_1,\ldots,\bm r_N)$ and the corresponding rows of each hidden
array. Write $\bm H^l\in\mathbb C^{N\times d_{\rm L}}$ for the array with
$i$th row $\bm h_i^l$. The query, key, and value maps are shared across
electron rows. Permuting the input therefore permutes their row labels. The
score in Equation~\eqref{eq:hmcva-score} carries one receiver and one source
label, so the same permutation relabels both indices. Row-wise softmax and the
sum in Equation~\eqref{eq:hmcva-aggregation} commute with this relabeling. The
output projection and residual perceptron are also shared across rows.
Consequently,
\begin{equation*}
\bm H^L(\Pi\bm R)=\Pi\bm H^L(\bm R).
\end{equation*}
The fixed functions $\chi_j(\bm r_i)$ are evaluated row by row and
transform in the same way. An electron permutation therefore only reorders the
rows of each orbital array
$[\phi_j^m(\bm r_i;\{\bm r_{/i}\})]_{i,j=1}^{N}$. Every determinant acquires
the same factor $\det(\Pi)$, and
\begin{equation*}
\Psi(\Pi\bm R)=\det(\Pi)\Psi(\bm R).
\end{equation*}
For an exchange of two electrons, $\det(\Pi)=-1$. The wavefunction is therefore fermionically antisymmetric.

The magnetic boundary condition follows independently. Translating
electron $i$ by $\bm L_a$ leaves its learned periodic coefficient unchanged.
The section $\chi_j$ multiplies every entry in row $i$ by the common transition
factor in Equation~\eqref{eq:section-transition}. Because the determinant is
linear in that row,
\begin{equation*}
\Psi(\ldots,\bm r_i+\bm L_a,\ldots)
=e^{\ii\Lambda_a(\bm r_i)}
 \Psi(\ldots,\bm r_i,\ldots).
\end{equation*}
This reproduces the many-body magnetic boundary condition in
Equation~\eqref{eq:magnetic-boundary}.

\section{Complex VMC optimization}
\label{app:complex-vmc}

During training, Metropolis--Hastings sampling uses ratios of
$|\Psi|^2$, so the wavefunction normalization is not needed.
For a symmetric proposal, the log probability ratio is
$2[u_{\bm\theta}(\bm R')-u_{\bm\theta}(\bm R)]$.
Evaluating this ratio in log amplitude avoids numerical overflow and underflow.
We derive the magnetic local energy and MinSR update below, and describe
the independent energy evaluation separately.

\subsection{Magnetic local energy}

In a magnetic field, the kinetic energy depends on both the amplitude
and spatial phase of a complex wavefunction. Locally away from nodes, write
\begin{equation*}
\Psi(\bm R)
=
e^{u(\bm R)}e^{\ii\phi(\bm R)},
\quad
u(\bm R)=\log|\Psi(\bm R)|.
\end{equation*}
For the kinetic operator in Equation~\eqref{eq:hamiltonian},
\begin{equation*}
\hat T
=
\frac{1}{2}\sum_i
\left(
-\ii\bm\nabla_i-\bm A_i
\right)^2,
\quad
\bm A_i=\bm A(\bm r_i),
\end{equation*}
where $\bm\nabla_i\cdot\bm A_i=0$ in the symmetric gauge. The kinetic
local energy is
\begin{equation}
\begin{aligned}
T_{\mathrm{loc}}(\bm R)
=
&-\frac{1}{2}\sum_i
\Big[
\nabla_i^2u
+|\bm\nabla_i u|^2
-|\bm\nabla_i\phi-\bm A_i|^2\\
&+\ii\big(
\nabla_i^2\phi
+2\bm\nabla_i u\cdot
(\bm\nabla_i\phi-\bm A_i)
\big)
\Big].
\end{aligned}
\label{eq:complex-kinetic}
\end{equation}
The phase gradient and vector potential enter together through
$\bm\nabla_i\phi-\bm A_i$. The Monte Carlo distribution depends only on
$|\Psi|^2=e^{2u}$, while the magnetic kinetic energy depends explicitly on the
phase through $\bm\nabla_i\phi-\bm A_i$. Adding the
moir\'e potential and Coulomb interaction gives the full local energy,
\begin{equation}
\begin{aligned}
E_{\mathrm{loc}}(\bm R)
=&
T_{\mathrm{loc}}(\bm R)
+\sum_iV_{\mathrm M}(\bm r_i)\\
&+\sum_{i<j}v_{\mathrm{Ew}}(\bm r_i-\bm r_j)
+\frac{N}{2}\xi_{\mathrm M}.
\end{aligned}
\label{eq:complete-local-energy}
\end{equation}

The local energy at configuration $\bm R$ is a pointwise ratio, not a
configuration-dependent eigenvalue. For an approximate complex trial state,
$[\hat H\Psi](\bm R)$ need not share the phase of $\Psi(\bm R)$. Consequently,
$E_{\mathrm{loc}}(\bm R)$ may be complex, while its probability-weighted
average is the real Rayleigh quotient in Equation~\eqref{eq:local-energy}
when the trial state belongs to the operator domain of the self-adjoint
Hamiltonian with the prescribed magnetic boundary condition. Under this
condition, the imaginary part cancels in the
exact expectation, although a finite Monte Carlo batch may retain a statistical
imaginary residual. For an exact eigenstate, the local energy equals the same real
eigenvalue at every non-nodal configuration \cite{foulkes2001qmc}. The
configuration-dependent imaginary part must still be retained because it
couples to the phase response in the energy gradient.

\subsection{Independent energy evaluation}

We evaluate the trained Psiformer states in
Section~\ref{sec:variational-benchmark} with the magnetic quadratic form,
\begin{equation*}
E=\left\langle
\frac12\sum_i\left[
|\bm\nabla_i u|^2+|\bm\nabla_i\phi-\bm A_i|^2
\right]+V(\bm R)
\right\rangle_{|\Psi|^2},
\end{equation*}
where $V$ contains the moir\'e potential and the complete periodic Coulomb
interaction. For a normalized trial state satisfying the magnetic
boundary condition, square-integrable first covariant derivatives in the weak
sense and a finite potential-energy integral suffice to define this variational
energy. A classical Laplacian at every configuration is not required. For
states in the operator domain, integration by parts recovers the local-energy
expectation. For nonsmooth trial states, derivative jumps can contribute
interface terms that a pointwise Laplacian omits. Statistical convergence of
the estimates must be assessed separately.

With the denominator regularization described after
Equation~\eqref{eq:radial-activation}, the implemented activation is
continuously differentiable in the real and imaginary
parts of its input, including at zero. Its second derivatives are generally
not continuous there. The regularity of the full ansatz also depends on the
absolute-value attention score, residual rescaling, and magnetic boundary
matching.

The conventional Complex-Psiformer local-energy samples have heavy
tails and show finite-sample differences from the quadratic-form estimates.
We use the same quadratic-form protocol for the independent Real--Complex
Psiformer comparisons. The conventional complex local energy remains the
training signal and an evaluation diagnostic.

Table~\ref{tab:hf-benchmark} retains the conventional local-energy summaries
after 10,000 training steps. The HF entries use Real-SlaterNet seed $1$ and
Complex-SlaterNet seed $0$. For the $18$-cell entries, the quoted uncertainties
are the local-energy sample standard deviations divided by $\sqrt{384}$.
They neglect correlations between samples and describe neither variation
across training seeds nor uncertainty in the converged HF minimum. The
HF-relative energy reductions use these conventional estimates throughout.
The independent quadratic-form evaluations provide the quantitative
Real--Complex Psiformer comparison.

\begin{figure}[t]
\centering
\includegraphics[width=0.9\columnwidth]{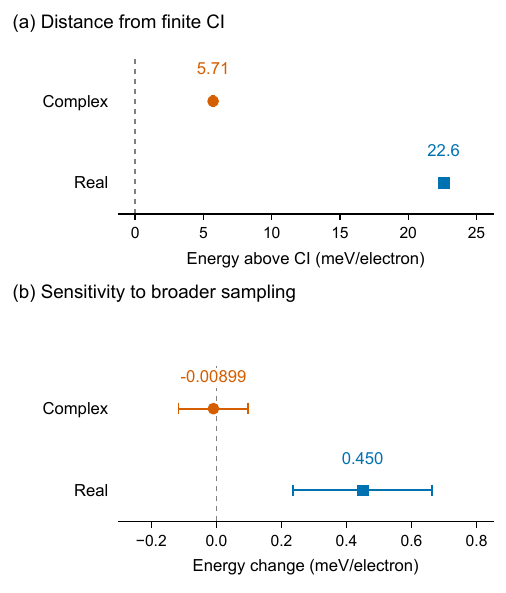}
\caption{Energy comparison and sampling sensitivity for the
$18$-cell, $N=2$, $n_\phi=2$ benchmark.
(a) Energy above the finite CI reference for the primary evaluation at
$\alpha=0.900$. Zero denotes the 40-orbital, 780-determinant CI energy.
(b) Change in the estimated energy when the same trained state is evaluated
independently at $\alpha=0.750$, relative to the primary estimate. The
dashed line denotes no change. Error bars show standard errors across 12
fixed walker groups, combined in quadrature in panel (b). Those in panel
(a) are smaller than the markers. Both states use training seed $1$ and
10,000 optimization steps. Each evaluation retains 393,216 configurations.}
\label{fig:n2-benchmark}
\end{figure}

Training uses configurations sampled from $|\Psi_{\bm\theta}|^2$.
After fixing the trained parameters, we
evaluate the state using a sampling density proportional to
$|\Psi|^{2\alpha}$. The exponent $0<\alpha<1$ increases the relative
probability of visiting low-amplitude regions, where kinetic-energy
estimators can fluctuate strongly. Ordinary sampling is recovered at
$\alpha=1$. For an observable $O$, we estimate its expectation under
$|\Psi|^2$ by $\sum_s w_s O(\bm R_s)/\sum_s w_s$, with
$w_s=|\Psi(\bm R_s)|^{2(1-\alpha)}$ for sampled configurations $\bm R_s$.
This reweighting leaves the target expectation unchanged in the sampling
limit. The exponent controls evaluation only and does not alter the trained
wavefunction.

The primary evaluation uses $\alpha=0.900$, 384 walkers, 20,000 burn-in
transitions, and 1,024 retained
sweeps separated by 20 Metropolis transitions, giving 393,216 retained
configurations per state. The network parameters, proposal scale, and estimator
settings remain fixed, and all retained configurations enter the estimates
without clipping. An independent calculation with $\alpha=0.750$ checks
sensitivity to broader sampling for the $N=2$ and $25$-cell benchmarks.

Sampling standard errors are estimated from 16 consecutive time blocks for the
$18$-cell, $N=12$ states and 12 fixed walker groups for $N=2$
and the $25$-cell, $N=12$ states. Each time block contains 64 retained
sweeps of all walkers, while each walker group contains 32 trajectories over
all retained sweeps. We recompute the self-normalized estimate within each
block or group. The sampling standard error is the sample standard deviation
of these estimates divided by the square root of their number.
For the $18$-cell seed-$1$ pair at $n_\phi=2$, the Complex and Real energies
are $-57.4$ and $-55.5\,\mathrm{meV}$ per electron, with standard errors of
$0.136$ and $0.0727\,\mathrm{meV}$ per electron. The corresponding
$N=2$ energies are $-29.0$ and $-12.1\,\mathrm{meV}$ per electron, with
standard errors of $0.0657$ and $0.165\,\mathrm{meV}$ per electron.
Adding the squared sampling standard errors and taking the square
root gives standard errors of $0.154$ and
$0.178\,\mathrm{meV}$ per electron for the respective energy differences.
These empirical errors describe sampling at fixed parameters. In the
$25$-cell comparison, the descriptive standard error of the difference
between the two five-seed means is $0.484\,\mathrm{meV}$ per electron,
obtained by adding the two sample variances divided by five and taking
the square root, with no covariance assigned between models.
Matching seed indices does not imply paired initial wavefunctions or
sampling trajectories. The observed spread of endpoint estimates also
contains residual Monte Carlo noise. Individual primary energies and their
sampling errors are supplied in \texttt{data/25cell\_seed\_energies.csv}.

For the $18$-cell, $N=12$ Complex and Real states, the importance-weight
effective sample sizes are $0.804$ and $0.782$ of the retained counts.
These fractions measure weight concentration and omit Markov-chain
autocorrelation. The first and second halves of the evaluations,
summarized by the means of their block estimates, differ by
$0.189$ and $0.150\,\mathrm{meV}$ per electron.
Doubling the block length to 128 retained sweeps gives standard
errors of $0.140$ and $0.0928\,\mathrm{meV}$ per electron for Complex and
Real, respectively. Additional block lengths from 1 to 256 sweeps are
reported in the accompanying data.

We also assess how strongly individual configurations affect these
energy estimates. On omitting one configuration at a time, the largest
observed changes are $0.114$ and $0.00491\,\mathrm{meV}$ per electron
for Complex and Real, respectively. These omissions serve only as sensitivity checks,
and all configurations remain in the reported estimates. The energy
ordering of this frozen pair survives these checks, while the errors remain
sensitive to blocking and rare sampled configurations. Longer independent
evaluations are needed to assess residual sampling bias and contributions
from unvisited regions with large spatial derivatives.

\begin{figure*}[t]
\centering
\includegraphics[width=\linewidth]{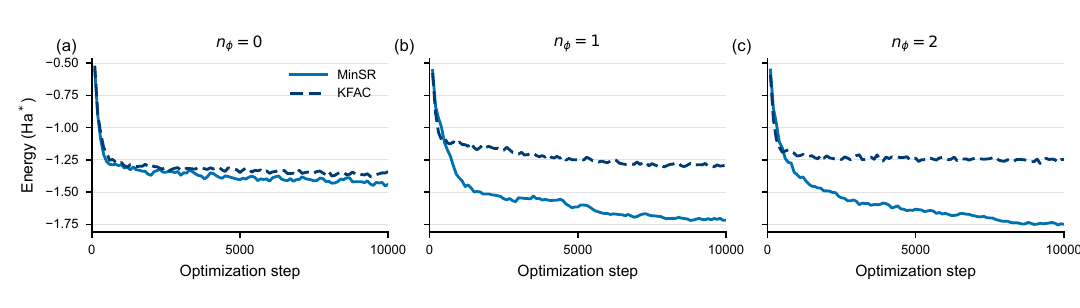}
\caption{Optimizer comparison for Real Psiformer. Panels (a)--(c) show the variational-energy convergence at $n_\phi=0$, $1$, and $2$, respectively. Solid curves show MinSR optimization and dashed curves show KFAC optimization.}
\label{fig:optimizer-comparison}
\end{figure*}

Figure~\ref{fig:n2-benchmark} separates the $N=2$ energy comparison from
its sampling check. The Complex and Real states lie $5.71$ and
$22.6\,\mathrm{meV}$ per electron above the 40-orbital, 780-determinant
CI reference. Changing the evaluation exponent from $0.900$ to $0.750$
shifts their energies by $-0.00899$ and $0.450\,\mathrm{meV}$ per electron,
respectively. The Complex shift is small relative to its combined sampling
error of $0.107\,\mathrm{meV}$ per electron. The Real shift is about
$2.11$ times its combined error of $0.213\,\mathrm{meV}$ per electron,
indicating residual sampling sensitivity. Both evaluations retain
the same energy ordering for this single training pair.
The plotted gaps use the fixed CI matrix energy as the reference.
Their error bars describe neural-state sampling and exclude the residual
error of the finite CI basis.

\subsection{Finite CI reference}
\label{app:ci-reference}

To construct the $N=2$ reference, we orthonormalize 448 torus Landau
functions spanning 224 Landau levels on a $96\times128$ midpoint grid.
We diagonalize the one-body Hamiltonian, including the moir\'e potential,
in this basis and retain its 40 lowest eigenstates. All 780 two-electron determinants
formed from these orbitals enter the CI calculation. Coulomb matrix elements
are assembled from Fourier transforms of orbital products, using supercell
reciprocal indices from $-16$ to $16$ and omitting the zero vector.
The Madelung self term is included once. The magnetic image cutoff is 12,
and the Ewald self term uses cutoff 8.

With the other cutoffs fixed, increasing the Landau-level count
from 192 to 224 changes the energy by $8.93\times10^{-4}\,\mathrm{meV}$ per
electron. Enlarging the active space from 36 to 40 orbitals changes it by
$3.90\times10^{-5}\,\mathrm{meV}$ per electron. These finite-cutoff changes
do not bound the remaining error relative to the continuum ground state.

An independent coordinate-space check of the same frozen CI state
uses 16 scrambled Sobol replicates, each with $2^{17}$ configurations.
The energy evaluated with the production Ewald interaction minus the CI
matrix energy is $(-2.30\pm2.09)\times10^{-4}\,\mathrm{meV}$ per electron.
The estimate uses the ratio of the pooled energy and normalization integrals.
Its empirical standard error is obtained across the randomized replicates,
retaining the covariance between numerator and denominator.

\subsection{Sample-space MinSR}

\begin{table*}[t]
\centering
\caption{Network and optimization settings for the Real and Complex SlaterNets.
Dimensions are stated over each model's number field, and parameter counts
give the trainable real scalar degrees of freedom of the neural HF references.}
\label{tab:slater_hyperparameters}
\small
\setlength{\tabcolsep}{3pt}
\renewcommand{\arraystretch}{1.08}
\begin{tabular}{@{}p{0.43\textwidth}p{0.23\textwidth}
p{0.27\textwidth}@{}}
\hline\hline
Parameter
& Real
& \shortstack[l]{Complex} \\
\hline

\multicolumn{3}{@{}l}{Architecture} \\
Hidden dimension $d_{\rm L}$
& $64\,(\mathbb R)$ & $32\,(\mathbb C)$ \\
Periodic-feature dimension $d_f$
& $4\,(\mathbb R)$ & $2\,(\mathbb C)$ \\
Number of Layers $L$
& $3$ & $3$ \\
Hidden activation
& $\tanh$ & signed radial $\tanh$ \\
Trainable real parameters (neural HF)
& $14{,}272$ & $13{,}600$ \\

\hline

\multicolumn{3}{@{}l}{Training} \\
Initial learning rate $\eta_0$
& $0.03$ & $0.03$ \\

\hline

\multicolumn{3}{@{}l}{Optimization} \\
Optimizer
& MinSR & MinSR \\
Damping
& $3\times10^{-2}$ & $3\times10^{-2}$ \\
Squared metric-norm constraint
& $5\times10^{-4}$ & $5\times10^{-4}$ \\
Score-Jacobian chunk
& $16$ samples & $16$ samples \\
\hline\hline
\end{tabular}

\end{table*}

\begin{table*}[t]
\centering
\caption{Model-specific settings for the representation-matched
Psiformer pair. Dimensions are stated over the number field used by each
model. Parameter counts refer to trainable real scalar degrees of freedom.}
\label{tab:psiformer_hyperparameters}
\small
\setlength{\tabcolsep}{3pt}
\renewcommand{\arraystretch}{1.08}
\begin{tabular}{@{}p{0.43\textwidth}p{0.23\textwidth}
p{0.27\textwidth}@{}}
\hline\hline
Parameter
& \shortstack[l]{Real}
& \shortstack[l]{Complex} \\
\hline

\multicolumn{3}{@{}l}{Architecture} \\
Hidden dimension $d_{\rm L}$
& $64\,(\mathbb R)$ & $32\,(\mathbb C)$ \\
Periodic-feature dimension $d_f$
& $4\,(\mathbb R)$ & $2\,(\mathbb C)$ \\
Attention heads $N_{\rm heads}$
& $2$ & $2$ \\
Query/key dimensions $d_q=d_k$
& $16$ & $8$ \\
Value dimension $d_v$
& $16$ & $8$ \\
Hidden activation
& $\tanh$ & signed radial $\tanh$ \\
Number of Layers $L$
& $3$ & $3$ \\
Number of determinant
& 4 & 4 \\
Trainable real parameters
& $43{,}456$ & $40{,}480$ \\

\hline

\multicolumn{3}{@{}l}{Training} \\
Initial learning rate $\eta_0$
& $0.03$ & $0.03$ \\
\hline

\multicolumn{3}{@{}l}{Optimization} \\
Optimizer
& MinSR & MinSR \\
Damping
& $3\times10^{-2}$ & $3\times10^{-2}$ \\
Squared metric-norm constraint
& $5\times10^{-4}$ & $5\times10^{-4}$ \\
Score-Jacobian chunk
& $16$ samples & $16$ samples \\

\hline\hline
\end{tabular}
\end{table*}

Wavefunction optimization must retain the phase dependence of the
magnetic local energy. For a batch $\{\bm R_b\}_{b=1}^{M}$, define the
centered local energy
\begin{equation*}
\widetilde E_b
=
E_{\mathrm{loc},\bm\theta}(\bm R_b)
-
\frac{1}{M}\sum_{b'=1}^{M}
E_{\mathrm{loc},\bm\theta}(\bm R_{b'}).
\end{equation*}
Using the amplitude and phase derivatives
$J^u_{b\mu}=\partial_{\theta_\mu}u_{\bm\theta}(\bm R_b)$ and
$J^\phi_{b\mu}=\partial_{\theta_\mu}\phi_{\bm\theta}(\bm R_b)$,
we obtain the Monte Carlo energy-gradient estimator under the
assumptions stated in Section~\ref{sec:minsr},
\begin{equation*}
g_\mu
=
\frac{2}{M}\sum_b
\left[
\widetilde J^u_{b\mu}\operatorname{Re}\widetilde E_b
+
\widetilde J^\phi_{b\mu}\operatorname{Im}\widetilde E_b
\right],
\end{equation*}
Here the tilde denotes subtraction of the batch mean. The amplitude
response couples to the real part of the local-energy fluctuation, while the
phase response couples to its imaginary part. Both terms contribute to the
variational energy gradient.

Stochastic reconfiguration measures updates through their effect on the
wavefunction. Retaining
both amplitude and phase responses gives the real SR metric
\cite{sorella1998stochastic},
\begin{equation}
S_{\mu\nu}
=
\frac{1}{M}\sum_b
\left(
\widetilde J^u_{b\mu}\widetilde J^u_{b\nu}
+
\widetilde J^\phi_{b\mu}\widetilde J^\phi_{b\nu}
\right).
\label{eq:sr-metric}
\end{equation}
Stacking the centered Jacobians into the matrix $X$ defined in
Equation~\eqref{eq:full-complex-jacobian}, we introduce
\begin{equation*}
\bm y
=
\frac{1}{\sqrt M}
\begin{pmatrix}
\operatorname{Re}\widetilde{\bm E}\\
\operatorname{Im}\widetilde{\bm E}
\end{pmatrix},
\qquad
\widetilde{\bm E}
=
\begin{pmatrix}
\widetilde E_1\\
\vdots\\
\widetilde E_M
\end{pmatrix}.
\end{equation*}
The metric and gradient then become
\begin{equation*}
S=X^{\mathsf T}X,
\qquad
\bm g=2X^{\mathsf T}\bm y.
\end{equation*}

A direct SR update requires a linear solve in the $P$-dimensional
parameter space. The push-through identity evaluates the same update in sample
space,
\begin{equation}
\begin{aligned}
\Delta\bm\theta
&=
-\eta
\left(
S+\lambda I_P
\right)^{-1}
\bm g\\
&=
-2\eta
X^{\mathsf T}
\left(
XX^{\mathsf T}+\lambda I_{2M}
\right)^{-1}
\bm y.
\end{aligned}
\label{eq:minsr-update-derived}
\end{equation}
This is the sample-space form used in MinSR
\cite{chen2024minsr,rende2024sridentity}. It reduces the linear solve from the
$P$-dimensional parameter space to the $2M$-dimensional sample space. The
parameters and linear system remain real. The complex wavefunction enters
through the amplitude and phase responses $J^u$ and $J^\phi$, together with
the real and imaginary parts of the centered local energy.

The push-through identity in
Equation~\eqref{eq:minsr-update-derived} is an algebraically exact rewrite of
the damped SR update for the sampled metric. It retains correlations among all real trainable parameters within the
finite batch, without factorizing the metric by layer. The $2M$-dimensional MinSR system includes
both amplitude and phase responses. All models in the main comparison use the
same implementation and numerical settings.

In training, the real and imaginary local energies are clipped
separately to three median absolute deviations about their respective batch
medians. The clipped components are then centered to construct $\bm y$.
After the MinSR solve, we rescale the update by a common factor if its squared
norm in the sampled SR metric exceeds the constraint listed in
Tables~\ref{tab:slater_hyperparameters} and \ref{tab:psiformer_hyperparameters}.
Clipping changes the energy-gradient estimator and is confined to the
optimization target. The independent quadratic-form evaluations retain all
sampled values without clipping.

\subsection{Auxiliary KFAC comparison}

KFAC approximates the parameter-space metric with layer-wise Kronecker
factors \cite{martens2015kfac}. This reduces the cost of storing and inverting
the full metric but neglects correlations between different layer blocks.
Ref.~\cite{geier2025attention} also used KFAC to optimize Real Psiformer. We
therefore compare KFAC and MinSR for the same Real Psiformer model at
$n_\phi=0,1,$ and $2$, as shown in
Figure~\ref{fig:optimizer-comparison}.

MinSR reaches a lower variational energy at $n_\phi=0$, even without
magnetic flux. The difference increases substantially at finite flux. At both
$n_\phi=1$ and $2$, the KFAC trajectories saturate at higher energies, while
MinSR continues to lower the energy. This comparison motivates the use of
MinSR in the main benchmark.

\begin{table*}[t]
\centering
\caption{Architectural, training, sampling, and numerical settings shared by
all four models.}
\label{tab:shared_hyperparameters}
\small
\setlength{\tabcolsep}{3pt}
\renewcommand{\arraystretch}{1.08}
\begin{tabular}{@{}p{0.14\textwidth}p{0.30\textwidth}
p{0.50\textwidth}@{}}
\hline\hline
Category & Parameter & Value \\
\hline
Training
& Training iterations
& $10{,}000$ \\
& Checkpoint and validation interval
& $250$ iterations \\
& Learning-rate schedule
& $\eta_t=\eta_0(1+t/t_0)^{-1}$ \\
& Learning-rate delay $t_0$
& $2000$ \\
& Local-energy clipping threshold
& $3$ MAD \\
& Real and complex precision
& \texttt{float64} and \texttt{complex128} \\
& Kinetic-energy chunk size
& $24$ samples \\

\hline

MCMC
& Batch size
& $384$ \\
& Burn-in iterations
& $0$ \\
& Initial proposal scale
& $0.5\,a_B^*$ \\
& Proposal-adaptation interval
& $50$ iterations \\
& Target acceptance ratio
& $0.5$ \\

\hline

Boundary and Ewald
& Ewald splitting length $\eta_{\mathrm{Ew}}$
& $\sqrt{\Omega/\pi}$ \\
& Ewald image indices
& $|n_1|,|n_2|\leq5$ \\
& Number of fixed magnetic functions
& $N=12$ \\
& Magnetic-function image indices
& $|n|\leq8$ \\
\hline\hline
\end{tabular}
\end{table*}

\section{Network and optimization hyperparameters}
\label{app:hyperparameters}

Calculations and derived comparisons use the unrounded values. Tables~\ref{tab:slater_hyperparameters},
\ref{tab:psiformer_hyperparameters}, and
\ref{tab:shared_hyperparameters} list the frozen settings for the main
four-model benchmark. The neural HF baselines use one determinant, while the
Psiformers use four. All models have three neural layers and share
the sampling settings and MinSR schedule listed in the tables.
The fixed magnetic functions contain no trainable
parameters and are excluded from the parameter counts.

We also performed supplementary experiments with four-determinant Real and
Complex SlaterNets, using the same hyperparameters as the neural HF references
in Table~\ref{tab:slater_hyperparameters}. The trainable real parameter counts
increase to $18{,}944$ and $15{,}904$, respectively. Their determinant values
are summed directly without independent mixing coefficients. These supplementary architectural controls have training diagnostics
separate from the frozen Real--Complex comparisons in
Section~\ref{sec:variational-benchmark}.

The Psiformer pair matches the hidden real dimension, depth, determinant count,
and attention widths. Complex Psiformer has slightly fewer trainable parameters.
Its attention branch also uses a warmup ramp and a cap on the
residual root-mean-square amplitude relative to the input, as specified in
the code configurations. The benchmark fixes these settings and the number
of optimization steps. Hyperparameter-search budgets and total GPU time are
not matched.

The independent density and structure-factor evaluation uses the $18$-cell
system at $n_{\phi}=0,\ldots,4$. The density profiles in
Figure~\ref{fig:density-comparison} and component estimates in
Figure~\ref{fig:structure-factor-components} use all 393,216 configurations
of each state. These estimates use separate Monte Carlo samples from the
momentum-space maps in Figure~\ref{fig:structure-factor-response}. All five
flux sectors use training seed $1$ and were optimized separately.

\section{Charge-order diagnostics}
\label{app:charge-order}

The density maps use the lattice vectors of Section~\ref{sec:system}.
The Cartesian coordinates follow from
$\bm r=(x,y)=u_1\bm L_1+u_2\bm L_2$, with $0\leq u_1,u_2<1$, so that
\begin{equation*}
\frac{x}{a_{\mathrm M}}=n_1u_1+\frac{n_2u_2}{2},\qquad
\frac{y}{a_{\mathrm M}}=\frac{\sqrt3\,n_2u_2}{2}.
\end{equation*}
The plotted origin is the cell corner, whereas $x_\parallel,x_\perp$ in the
magnetic-section construction are measured from the cell center $\bm l$.
The Jacobian multiplies $\rho$ and its spatial mean
$\langle\rho\rangle=N/\Omega$ by the same factor, leaving the plotted ratio
unchanged. We use the six primitive Brillouin-zone corners to probe
three-sublattice order,
\begin{equation}
\left\{\pm\frac{2\bm g_1+\bm g_2}{3},\;
\pm\frac{\bm g_1+2\bm g_2}{3},\;
\pm\frac{\bm g_1-\bm g_2}{3}\right\}.
\label{eq:moire-K-star}
\end{equation}

\subsection{Structure-factor components}

We evaluate the decomposition in
Equation~\eqref{eq:structure-factor-decomposition} at each reciprocal vector
before averaging over the six vectors in Equation~\eqref{eq:moire-K-star}.
Figure~\ref{fig:structure-factor-components} shows the resulting $S(K)$,
$S_{\rm disc}(K)$, and $S_{\rm conn}(K)$. At zero field, Complex and Real
Psiformer give $S(K)=0.707\pm0.002$ and $0.633\pm0.001$,
respectively. Uncertainties are time-block sampling standard errors.
The reciprocal-grid differences $\Delta S_{n_\phi}(\bm q)$ share
the same zero-field estimate. Comparisons between fluxes must retain the
covariance from this shared reference.

Complex Psiformer has its largest $S(K)$ at $n_\phi=2$, with
$S(K)=1.65\pm0.003$, compared with $0.851\pm0.002$ for Real
Psiformer. Their disconnected contributions are $1.07\pm0.003$ and
$0.301\pm0.001$, whereas the connected parts are
$0.583\pm0.001$ and $0.549\pm0.001$. The mean-density contribution
accounts for $95.8\%$ of the Complex--Real difference in $S(K)$ at this flux.
Real Psiformer has comparable total maxima near $n_\phi=1$ and $4$, again
with substantial disconnected contributions. The connected Complex values
range from $0.550$ to $0.661$, while the Real values range from $0.549$
to $0.627$, so the larger total structure factor does not imply larger
connected fluctuations.
At $n_\phi=4$, the Complex disconnected contribution is only
$S_{\rm disc}(K)=0.00136\pm0.000120$, consistent with the suppressed
three-sublattice mean-density pattern. These quantities probe the $K$ star
and leave open possible order at other wavevectors or order hidden by
symmetry averaging.

For example, equal statistical averaging over the three translated
honeycomb charge patterns cancels their mean-density Fourier component at
the $K$ star. Translation changes the phase of this component while leaving
its squared magnitude unchanged. The corresponding contribution to $S(K)$
then enters $S_{\rm conn}(K)$. This illustrates how crystalline correlations
can survive the loss of a mean-density motif. The present scan does not
establish that such averaging occurs in the high-field state.

\subsection{Three-sublattice occupation}

To compare the $18$-cell densities with the honeycomb motif, we assign every
electron to its nearest moir\'e-potential minimum using the periodic
Cartesian distance. The resulting mean well occupation is denoted by $n_\ell$,
with $\sum_{\ell=1}^{18}n_\ell=12$. For primitive vectors separated by
$60^\circ$, a well at $i\bm a_1+j\bm a_2$ belongs to sublattice
$s=(i-j)\bmod3$. We label $s=0,1,2$ by $A,B,C$, respectively. Each sublattice
contains six wells, and every nearest-neighbor bond connects different
sublattices. Let $\bar n_s$ be the mean within sublattice $s$ and
$\bar n=2/3$ the mean over all wells.

We quantify the sublattice pattern through the fraction of the well-to-well
occupation variation accounted for by the three sublattice means,
\begin{equation*}
f_3=\frac{6\sum_{s=A,B,C}(\bar n_s-\bar n)^2}
{\sum_{\ell=1}^{18}(n_\ell-\bar n)^2}.
\end{equation*}
For nonzero total variation, $f_3$ ranges from zero to one. It measures
three-sublattice modulation but does not specify which sublattices are occupied.
We therefore also report $d_{\rm h}$, the root-mean-square deviation of the
18 well occupations from an ideal honeycomb pattern with two occupied
sublattices and one empty sublattice. We take the smallest deviation among
the three choices of empty sublattice, without rearranging individual wells.
Quantum
fluctuations and finite localization can yield noninteger occupations even
in an ordered state.

\begin{table}[t]
\centering
\caption{Three-sublattice descriptors of all five independently
optimized $18$-cell flux sectors.
Larger $f_3$ indicates that sublattice means explain more of the spatial
variation, whereas smaller $d_{\rm h}$ indicates proximity to the ideal
honeycomb occupation motif. The deviation $d_{\rm h}$ is in electrons per
well.}
\label{tab:sublattice-flux}
\small
\begin{tabular}{ccccc}
\hline\hline
$n_\phi$ & \multicolumn{2}{c}{$f_3$} & \multicolumn{2}{c}{$d_{\rm h}$} \\
 & Complex & Real & Complex & Real \\
\hline
0 & 0.157 & 0.000819 & 0.522 & 0.540 \\
1 & 0.534 & 0.454 & 0.335 & 0.352 \\
2 & 0.969 & 0.285 & 0.131 & 0.484 \\
3 & 0.144 & 0.118 & 0.455 & 0.461 \\
4 & 0.00316 & 0.448 & 0.511 & 0.351 \\
\hline\hline
\end{tabular}
\end{table}

Table~\ref{tab:sublattice-flux} shows that the Complex state is closest to
the ideal honeycomb pattern at $n_\phi=2$, where its sublattice means account
for $96.9\%$ of the occupation variation. At zero field, its sublattice means
are $(0.577,0.861,0.562)$, without the depleted sublattice characteristic of
the honeycomb motif. A larger $S(K)$ alone therefore does not identify that
pattern. At $n_\phi=4$, the Real state is closer to the ideal motif, so the
relative agreement at $n_\phi=2$ does not extend across the full scan.

These occupation measures were chosen after examining the flux scan and
use the same 393,216 weighted configurations as the density profiles.
We report point estimates because uncertainties in $f_3$ and $d_{\rm h}$
require covariances between well occupations, which are unavailable from
the archived per-well standard errors.

\subsection{Seed dependence in the $25$-cell system}

\begin{figure}[t]
\centering
\includegraphics[width=0.92\columnwidth]{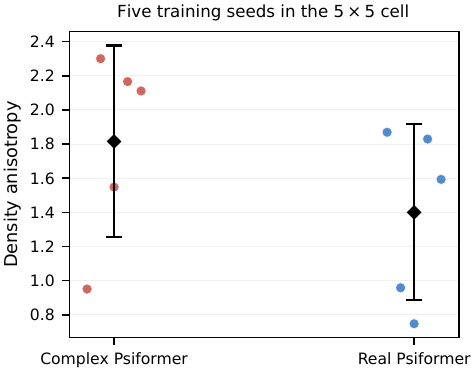}
\caption{Training-seed dependence of the $25$-cell density
anisotropy, defined as the directional range divided by the mean.
Points show all five independently optimized states of each
Psiformer at $N=12$ and $n_\phi=2$. Diamonds and bars show the mean and sample
standard deviation over seeds. The $5\times5$ cell is not exactly commensurate
with the candidate $K$ or $M$ ordering vectors, so this is a finite-cell
measure of directional variation.}
\label{fig:seed-anisotropy}
\end{figure}

For the incommensurate $5\times5$ geometry, we evaluate density anisotropy
at the allowed reciprocal-grid points nearest to the three independent
$M$ directions. We take the largest minus the smallest of their
disconnected structure-factor values and divide by their mean. This
dimensionless ratio measures the variation among the three directions.
Across the five independently optimized seeds, its mean and sample standard
deviation are $1.82\pm0.562$ for Complex Psiformer and $1.40\pm0.515$ for
Real Psiformer. Figure~\ref{fig:seed-anisotropy} shows all individual values.
The broad overlap shows that the apparent anisotropy depends on initialization
and does not consistently separate the two architectures. The filling
$12/25$ and allowed reciprocal vectors also differ from the half-filled
stripe reference of Ref.~\cite{li2021imagingwigner}.

\FloatBarrier
\end{appendix}

\bibliography{ref}

\end{document}